\documentclass[preprint,5p,times,twocolumn,num,sort&compress,nopreprintline]{elsarticle}  % final：终稿。5p：紧凑双栏。times：字体 times new roman。twocolumn：双栏

\usepackage{amsmath}
\usepackage{amssymb}   % for math
\usepackage{graphicx}
\usepackage{float}
\usepackage{enumitem}
\usepackage[colorlinks=true, linkcolor=blue, citecolor=blue, urlcolor=blue, pdfborder={0 0 0}]{hyperref}
\usepackage{setspace}
\usepackage{braket}
\usepackage{ulem}     % for delete line in text
\usepackage{xcolor}                          % 引入颜色宏包
\usepackage{listings}
\usepackage[switch]{lineno} % 引入宏包,对生成的 pdf 所有行编号，switch 选项保证左(右)侧行号在左(右)侧页边空白

\usepackage{array}
\usepackage{tabularx}

\usepackage{physics}

\allowdisplaybreaks[2]  % 允许公式在4级及以上的情况下分页

\usepackage{fancyhdr}
\fancypagestyle{firstpage}{
  \fancyfoot[L]{
    \footnotesize
    \textsuperscript{\dag}These authors contributed equally. \\
    \textsuperscript{*}Corresponding authors. Emails: kongxianghuaphysics@szu.edu.cn (X.K.), wji@ruc.edu.cn (W.J.)
  }
}

\begin{document}
% \linenumbers % 开始行号

\begin{frontmatter}

%% Title, authors and addresses

%% use the tnoteref command within \title for footnotes;
%% use the tnotetext command for theassociated footnote;
%% use the fnref command within \author or \address for footnotes;
%% use the fntext command for theassociated footnote;
%% use the corref command within \author for corresponding author footnotes;
%% use the cortext command for theassociated footnote;
%% use the ead command for the email address,
%% and the form \ead[url] for the home page:
%% \title{Title\tnoteref{label1}}
%% \tnotetext[label1]{}
%% \author{Name\corref{cor1}\fnref{label2}}
%% \ead{email address}
%% \ead[url]{home page}
%% \fntext[label2]{}
%% \cortext[cor1]{}
%% \affiliation{organization={},
%%             addressline={},
%%             city={},
%%             postcode={},
%%             state={},
%%             country={}}
%% \fntext[label3]{}

%%\title{Moir\'{e} Crystallography: Universal Geometric Decoding of Strained Twisted Bilayers from Real/Reciprocal-Space Imaging}

%%\title{Generalized Moir\'{e} Crystallography for Arbitrary Bilayers: Decoding Twist, Strain, and Primitive Coincident Cells from Imaging}

\title{Primitive-cell-resolved Crystallography for Moir\'{e} Bilayers from Imaging}

% 作者信息
\author[inst1]{Zhidan Li}                   
\author[inst1]{Xianghua Kong\corref{cor1}}  % 通讯作者
% 共同第一作者标注
%\fntext[fn1]{These authors contributed equally to this work.}
% 通讯作者标注
\cortext[cor1]{Corresponding author. Email: kongxianghuaphysics@szu.edu.cn (X.K.)}
%\cortext[cor2]{Corresponding author. Email: wji@ruc.edu.cn (W.J.)}
% 单位信息
\affiliation[inst1]{organization={College of Physics and Optoelectronic Engineering}, 
            addressline={Shenzhen University}, 
            city={ Shenzhen},
            postcode={518061},
            country={China}}
%\affiliation[inst2]{organization={Beijing Key Laboratory of Optoelectronic Functional Materials \& Micro-nano Devices, 
%            Department of Physics},
%            addressline={Renmin University of China}, 
%            city={ Beijing},
%            postcode={100872},
%            country={China}}
%\affiliation[inst3]{organization={Key Laboratory of Quantum State Construction and Manipulation (Ministry of Education)},
%            addressline={Renmin University of China}, 
%            city={ Beijing},
%            postcode={100872},
%            country={China}}

\begin{abstract}
Accurate geometric decoding of moir\'{e} bilayers from imaging is essential for engineering quantum systems. Existing schemes, limited by identity or aligned assumptions requiring diagonal beating-to-moiré transformations, do not apply to general non-aligned geometries and become underdetermined when buried layers are unresolved. We establish a primitive-cell-resolved moir\'{e} crystallography framework that treats the beating-to-moir\'{e} relation in full generality and introduces a complete descriptor set $\{\theta_r,\boldsymbol{\varepsilon},(T_{Mt},T_{Mb}),N_B\}$, where the integer moir\'{e}--layer matrices  $(T_{Mt},T_{Mb})$ and the beating number $N_B$ determine the commensurate unit cell. A hybrid analytical--numerical workflow reconstructs buried-layer lattices, solves Diophantine constraints to obtain $(T_{Mt},T_{Mb})$ and $N_B$, and extracts $(\theta_r,\varepsilon_b,\theta_u,\varepsilon_u)$ with Poisson effects and tensile/compressive branches treated on equal footing. Reanalyzing twisted bilayer graphene, we identify a $N_B=3$ primitive cell rather than a $N_B=9$ aligned supercell, reducing the atomistic basis threefold and correcting the moir\'{e} Brillouin-zone construction. The framework provides a crystallographically consistent route from imaging to primitive-cell-resolved atomistic and many-body models.
\end{abstract}

% %%Graphical abstract
% \begin{graphicalabstract}
% \includegraphics{grabs}
% \end{graphicalabstract}

% %%Research highlights
% \begin{highlights}
% \item Research highlight 1
% \item Research highlight 2
% \end{highlights}

\begin{keyword}
%% keywords here, in the form: keyword \sep keyword
Moir\'{e} Crystallography \sep Beating-Moir\'{e} Transformation \sep Primitive Moir\'{e} Cells\sep Beating Number
%% PACS codes here, in the form: \PACS code \sep code
%% \PACS 0000 \sep 1111
%% MSC codes here, in the form: \MSC code \sep code
%% or \MSC[2008] code \sep code (2000 is the default)
%% \MSC 0000 \sep 1111
\end{keyword}
\end{frontmatter}

\section{Introduction}
\label{sec:Intro}
Bilayer Moir\'{e} superlattices, exemplified by magic-angle twisted bilayer graphene (MATBG)~\cite{Ref_TBG_Theory_2007, Ref_TBG_Theory_2010_Tight_Binding_Model, Ref_TBG_Theory_2012, Ref_TBG_Theory_2011_BM_Model}, have advanced condensed matter physics by enabling correlated insulating and superconducting states~\cite{Ref_Cao_2018_Mott, Ref_Cao_2018_SC, Ref_SC_Science_2019, Ref_topo_SC_PRL_Balents_2018, Ref_Mott_SC_Senthil_PRX_2018}, topological phases~\cite{Ref_TBG_QHE_PRL_2011, Ref_TBG_QHE_PRL_2012, Ref_TBG_QHE_PRB_2012, Ref_TBG_QHE_Science_2018, Ref_TBG_QHE_NC_2023, Ref_topo_SC_PRL_Balents_2018, Ref_TBG_topology_PRL_Bernevig_2019, Ref_TBG_topology_PRL_MacDonald_2019}, and
moir\'{e} excitons~\cite{Ref_Moiron_JPCM_2012, Ref_Moiron_PRB_2013, Ref_Moiron_Nature_00_2019, Ref_Moiron_Nature_01_2019, Ref_Moiron_Nature_02_2019}. These phenomena arise because the moir\'{e} geometry, determined by the interlayer twist angle, heterostrain, and stacking registry, reconstructs the electronic structure and renormalizes the effective interactions at the moir\'{e} length scale. For instance, MATBG exhibits correlated phases and superconductivity precisely near the ``magic angle"~\cite{Ref_Cao_2018_Mott, Ref_Cao_2018_SC}, where the twist angle controls bandwidth and band topology. Strain provides another powerful tool for modulating electronic structures, tuning band flatness, shifting critical angles governing quantum phase transitions, and inducing pseudo-magnetic fields through lattice deformation~\cite{Ref_strain_FaradayDiscuss_2014, Ref_strain_PRB_2019, Ref_strain_PRL_2018, Ref_SC_Science_2019, Ref_strain_PRL_2021}. Beyond graphene-based systems, incorporating transition-metal dichalcogenides (TMDCs), two-dimensional (2D) magnets, and other layered 2D structures~\cite{Ref_cupratemoire_2023_NC, Ref_hBN_2021_NC, Ref_hBN_2024_NC, Ref_BlackP_2021_NE, Ref_BlackP_2024_NC} substantially broadens the moir\'{e} materials landscape, introducing new quantum phenomena driven by pronounced $d$-orbital physics~\cite{Ref_2019_TMDC_00, Ref_Moiron_Nature_00_2019, Ref_Moiron_Nature_01_2019, Ref_Moiron_Nature_02_2019, Ref_TBG_topology_PRL_MacDonald_2019}, robust excitonic effects~\cite{Ref_Moiron_Nature_00_2019, Ref_Moiron_Nature_01_2019, Ref_Moiron_Nature_02_2019, Ref_2019_TMDC_00, Ref_PFM_01, Ref_exciton_2025_PNAS, Ref_exciton_2023_PRL}, and emergent magnetic orders~\cite{Ref_exciton_2025_PNAS,Ref_MoireMagnet_00,Ref_MoireMagnet_01,Ref_MoireMagnet_02,Ref_MoireMagnet_03}. However, increased intralayer thickness in diverse moir\'{e} materials often obscures atomic-scale details of buried layers, hindering accurate structural determination essential for understanding fundamental physics and designing novel moir\'{e} quantum systems.

The decoding of moir\'{e} geometric parameters from real-space imaging and reciprocal-space diffraction data is fundamental for characterizing moir\'{e} quantum materials. However, current methodologies are largely limited to bilayer systems with thin top layers, where techniques such as scanning tunneling microscopy (STM)~\cite{Ref_uni_bi_axial_strain_model_2016, Ref_strain_PRL_2018, Ref_STM_01, Ref_STM_02, Ref_STM_03, Ref_STM_04, Ref_STM_05}, atomic force microscopy (AFM)~\cite{Ref_cAFM_01, Ref_cAFM_02}, and transmission electron microscopy (TEM)~\cite{Ref_TEM_01, Ref_TEM_02, Ref_TEM_03, Ref_TEM_04} permit atomic-resolution imaging of both constituent layers. Within this imaging regime, two models for moir\'{e} geometry determination are commonly used: (1) The uniaxial-strain model~\cite{Ref_Correlation_uniaxial_strain_model_2019, Ref_Correlation_uniaxial_strain_model_2020}, which assumes zero biaxial strain and extracts the twist angle ($\theta_r$), the uniaxial-strain direction ($\theta_u$), and the magnitude ($\varepsilon_u$) by mapping the three nearest local density of states (LDOS) maxima (representing beating lattice vectors) onto the moir\'{e} primitive vectors. This approach operates in the identity limit ($N_B = 1$), meaning the beating-to-moir\'{e} transformation is fixed as the identity matrix, effectively defining an identical model. While effective for small strain without biaxial effects, it neglects possible stacking-subtype modulations and provides constrained solutions for the moir\'{e}--layer transformation matrices ($T_{Mt}, T_{Mb}$), which can limit the accuracy of atomistic reconstruction. (2) The general-strain model under the small-angle approximation~\cite{Ref_uni_bi_axial_strain_model_2016, Ref_strain_PRL_2018}, which incorporates both uniaxial and biaxial strains. By enforcing alignment between the moir\'{e} and beating lattice vectors ($\mathbf{k}_{B\alpha} \parallel \mathbf{k}_{M\alpha}$), this aligned model restricts the beating-to-moir\'{e} transformation to a diagonal matrix to determine $T_{Mt}$ and $T_{Mb}$ and specify the moir\'{e} primitive cell. However, such alignment is not a crystallographic necessity. Fundamentally, the orientations and magnitudes of $\mathbf{k}_{B\alpha}$ and $\mathbf{k}_{M\alpha}$ are linked by a comprehensive $2 \times 2$ matrix that encompasses identity, diagonal, and non-diagonal cases. Furthermore, reported solutions predominantly describe tensile strain, with compressive cases rarely captured.

Despite these advances, imaging-based decoding that relies on the identical and aligned models remains constrained in three respects. First, many platforms currently driving the field, such as TMDCs and 2D magnets with appreciable intralayer thickness often fall outside their scope because the buried layer is not atomically resolved with standard STM/AFM/TEM, rendering the geometric decoding underdetermined. Second, even when both layers are resolved, these prevailing prescriptions do not guarantee identification of the primitive moir\'{e} cell: enforcing alignment between moir\'{e} and beating vectors, or adopting the identity limit can yield non-primitive supercells in non-aligned geometries. Determining the authentic primitive cell therefore requires a primitive-consistent moir\'{e}--layer matrices $(T_{Mt},T_{Mb})$ together with the beating number $N_B$. This information fixes the minimal periodicity and atomic basis and thereby enables consistent reciprocal-space indexing and band folding, underpins the correct moir\'{e} Brillouin-zone construction. These are prerequisites for locating van Hove singularities and interpreting the associated density-of-states features in correlated-regime findings. A primitive-consistent description also clarifies the organization of electron--lattice coupling in real space and supplies the correct simulation cell for atomistic calculations. As moir\'{e} platforms continue to diversify toward strongly correlated regimes, a rigorous expansion of the geometric descriptor set beyond twist and strain alone becomes necessary. Third, the small-angle and weak-strain approximations further restrict applicability to arbitrary twist and general strain states. These considerations motivate the framework developed below for decoding moir\'{e} geometry from imaging.

In this work, we introduce a generalized moir\'{e} crystallography framework that explicitly incorporates the beating number (\(N_B\)) and the moir\'{e}--layer transformation matrices (\(T_{Mt}, T_{Mb}\)) as fundamental geometric descriptors. These parameters complement established parameters such as the twist angle (\(\theta\)) and the strain tensor (\(\boldsymbol{\varepsilon}\)). Leveraging only top-layer features and the beating pattern extracted from real- and reciprocal-space images, the framework utilizes their consistency relations to reconstruct the bottom-layer lattice vectors and associated transformation matrices even when the bottom layer is not atomically resolved. This addresses a key gap in current methodologies, which have either equated beating periodicity directly with the moir\'{e} periodicity or relied on enhanced imaging penetration and resolution, thereby increasing experimental complexity and limiting broader applicability. Our framework also dispenses with alignment between the moir\'{e} and beating lattice vectors. By combining analytical relations with a numerical solver, it decodes the complete geometric parameter set \(\{\theta, \boldsymbol{\varepsilon}, (T_{Mt}, T_{Mb}), N_B\}\), with \(\boldsymbol{\varepsilon}\) covering uniaxial and biaxial as well as tensile, compressive, and mixed cases, without small-angle or weak-strain approximations. The methodology naturally admits multiple candidate solutions; physically relevant ones are identified using energy-minimization criteria or complementary experimental constraints. Applied to published STM data on twisted bilayer graphene from Ref.~\cite{Ref_strain_PRL_2018}, the framework identifies a primitive moir\'{e} cell about one third of the previously assumed size and reduces the number of atoms required for simulations by approximately 66\%, thereby correcting the structural model and lowering computational cost. Overall, the framework links precise geometric decoding with predictive design in moir\'{e} quantum materials.

\begin{figure*}[!t]
  \centering{
  \includegraphics[width=0.8\textwidth,trim=0.5mm 0.5mm 0.9mm 0.5mm,clip]{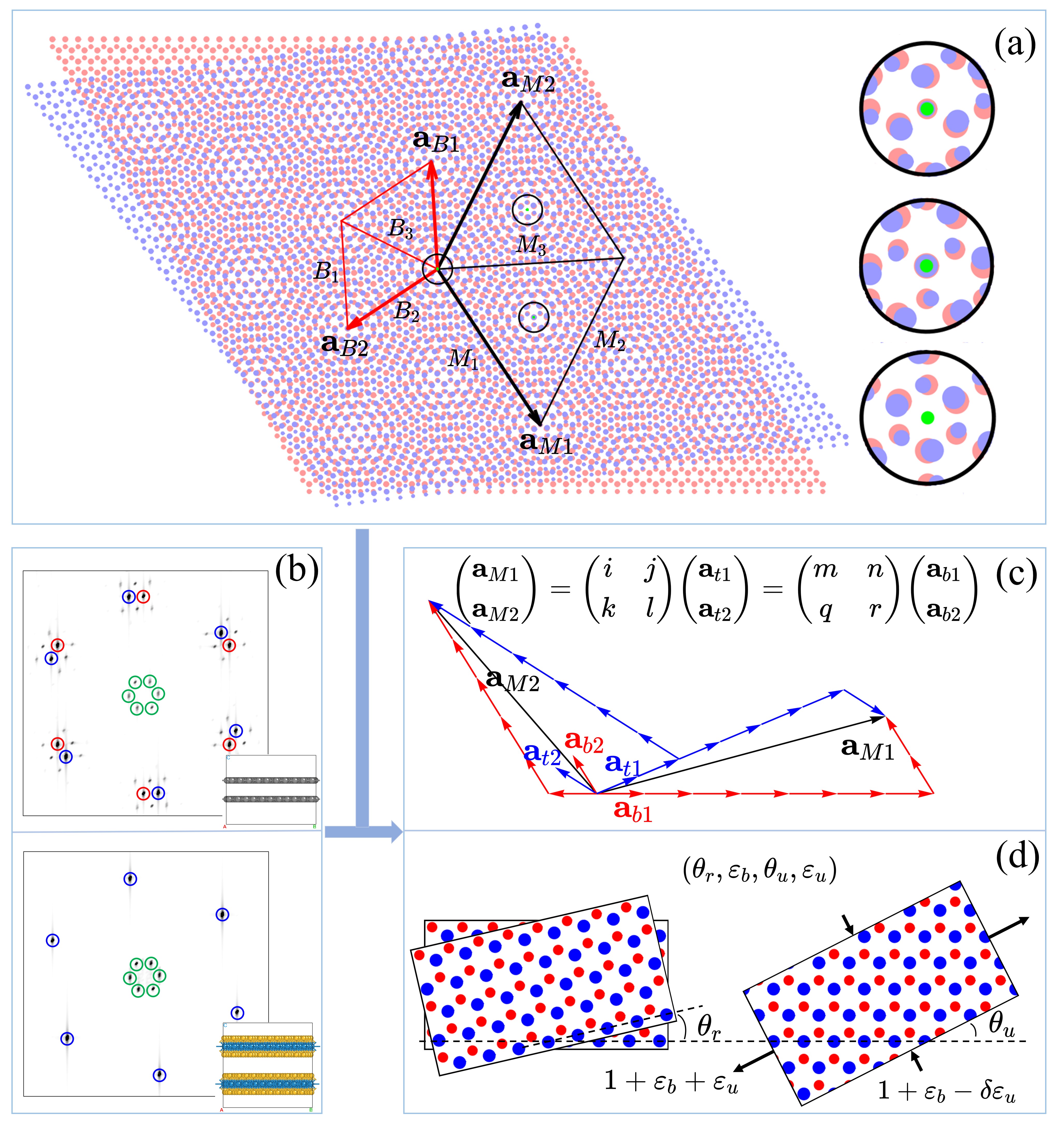}
  }
 \caption{(a) Schematic of the moir\'{e} (M) and beating (B) lattices for a bilayer with twist angle $6.836^{\circ}$ and zero strain. The primitive vectors are $\{\mathbf{a}_{M1},\mathbf{a}_{M2}\}$ and $\{\mathbf{a}_{B1},\mathbf{a}_{B2}\}$. The quantities $(M_1,M_2,M_3)=(|\mathbf{a}_{M1}|,|\mathbf{a}_{M2}|,|\mathbf{a}_{M1}+\mathbf{a}_{M2}|)$ and $(B_1,B_2,B_3)=(|\mathbf{a}_{B1}|,|\mathbf{a}_{B2}|,|\mathbf{a}_{B1}+\mathbf{a}_{B2}|)$ are defined accordingly. The three panels on the right enlarge the circled regions, showing registries at AA, BB, and hollow-hollow sites; green dots mark beating centers. (b) Reciprocal-space diffraction schematics. The top (bottom) panel illustrates cases where bottom-layer information is detectable (undetectable). Red, blue, and green circles indicate diffraction peaks from the top layer, bottom layer, and beating, respectively. Insets in the lower right corners indicate a thin top layer (top panel) and a thick top layer (bottom panel). (c) The moir\'{e} primitive vectors $\{\mathbf{a}_{M1},\mathbf{a}_{M2}\}$ are related to the top-layer lattice vectors $\{\mathbf{a}_{t1},\mathbf{a}_{t2}\}$ via an integer transformation matrix $T_{Mt}$ (or to the bottom-layer vectors $\{\mathbf{a}_{b1},\mathbf{a}_{b2}\}$ via $T_{Mb}$). (d) Key geometric control parameters for generating a moir\'{e} superlattice: twist angle $(\theta_r)$ and strain (biaxial $(\epsilon_b)$; uniaxial magnitude $(\epsilon_u)$ with direction $(\theta_u)$). The material's elastic response is characterized by its Poisson's ratio $(\delta)$. }
 \label{fig_01}
\end{figure*}

\section{Framework for moir\'{e} crystallography}
\label{sec:framework}

Two stacked 2D lattices, when subjected to a twist angle or strain, give rise to a long-period moir\'{e} superlattice structure. In real-space imaging, this structure manifests as a periodic modulation of the local stacking order at specific local registries, as shown in Fig.~\ref{fig_01}(a). However, these visually similar regions can possess distinct atomic-scale registries. The right panels of Fig.~\ref{fig_01}(a), for instance, show that the atomic-center registries include AA, BB, and hollow--hollow (HH). To formalize this distinction, the general, visually periodic pattern arising from these modulations is defined as the ``beating'' lattice, described by the primitive vectors $\{\mathbf{a}_{B1}, \mathbf{a}_{B2}\}$. The ``moir\'{e}'' lattice is defined as the Bravais lattice whose translations exactly repeat the full atomic configuration, described by $\{\mathbf{a}_{M1}, \mathbf{a}_{M2}\}$. The relationship between these two lattices is quantified by the beating number $N_B$, which counts the number of beating centers contained in one moir\'{e} primitive cell (see Eq.~\eqref{Eq_NB_def}). For the unstrained system with a $6.836^{\circ}$ twist angle shown in Fig.~\ref{fig_01}(a), there are three such centers per moir\'{e} cell, i.e., $N_B = 3$. The two lattices coincide only in the special case where $N_B = 1$, in which case all AA-type contrast maxima are atomically indistinguishable.

The fundamental distinction between the beating and moir\'{e} lattices is also manifest in reciprocal space. A typical diffraction pattern, schematized in Fig.~\ref{fig_01}(b) (top panel), exhibits three main features: (i) diffraction peaks from the top layer ($\{\mathbf{k}_{t1},\mathbf{k}_{t2}\}$, blue circles), (ii) peaks from the bottom layer ($\{\mathbf{k}_{b1},\mathbf{k}_{b2}\}$, red circles), and (iii) a set of six peaks surrounding the origin ($\Gamma$ point) that arise from the difference between the layer reciprocal vectors ($\{\mathbf{k}_{B1},\mathbf{k}_{B2}\}$, green circles). These innermost peaks define the reciprocal lattice of the beating pattern. They reflect the dominant real-space periodicity but are not, in general, the reciprocal vectors of the moir\'{e} superlattice, $\{\mathbf{k}_{M1},\mathbf{k}_{M2}\}$. In systems with thick layers such as TMDCs, the signal from the buried bottom layer may be below detection thresholds [Fig.~\ref{fig_01}(b), bottom panel], so only the top-layer and beating peaks are accessible. The following sections will establish the mathematical framework required to decode the complete geometric parameter set from this limited data.

\medskip
%\noindent\textbf{Real-space layer--moir\'{e} relation.}
\noindent\textbf{Real-space Moir\'{e}--Layer Transformation.}
%\noindent\textbf{The Moir\'{e} Superlattice and its Integer-Matrix Representation.}
As shown in Fig.~\ref{fig_01}(c), the real-space vectors of the moir\'{e} primitive cell $\{\mathbf{a}_{M1}, \mathbf{a}_{M2}\}$ are integer linear combinations of the primitive vectors of either the top ($\{\mathbf{a}_{t1},\mathbf{a}_{t2}\}$) or bottom ($\{\mathbf{a}_{b1}, \mathbf{a}_{b2}\}$) layer ~\cite{Ref_uni_bi_axial_strain_model_2016},
\begin{align}\label{Eq_aMatab}
    \left(
        \begin{array}{c}
            \mathbf{a}_{M1}\\
            \mathbf{a}_{M2}
        \end{array}
    \right)
    = T_{Mt}
    \left(
        \begin{array}{c}
            \mathbf{a}_{t1}\\
            \mathbf{a}_{t2}
        \end{array}
    \right)
    = T_{Mb}
    \left(
        \begin{array}{c}
            \mathbf{a}_{b1}\\
            \mathbf{a}_{b2}
        \end{array}
    \right),
\end{align}
with the integer transformation matrices being
\begin{align}
    T_{Mt} =
    \left(
        \begin{array}{cc}
            i & j \\
            k & l
        \end{array}
    \right)
    , \quad
    T_{Mb} =
    \left(
        \begin{array}{cc}
            m & n \\
            q & r
        \end{array}
    \right), \nonumber
\end{align}
where $i,j,k,l,m,n,q,$ and $r$ are integers. If these eight integers together with the primitives of either layer are known, the crystallography of the moir\'{e} lattice is fully determined.

\medskip
\noindent\textbf{Reciprocal-space Beating--Moir\'{e} Transformation.} The reciprocal beating vectors, $\{\mathbf{k}_{B1},\mathbf{k}_{B2}\}$, are fundamentally defined as the difference between the layer reciprocal vectors:
\begin{align}\label{Eq_Beating_def_fundamental}
    \left(
        \begin{array}{c}
            \mathbf{k}_{B1}\\
            \mathbf{k}_{B2}
        \end{array}
    \right)
    \equiv
    \left(
        \begin{array}{c}
            \mathbf{k}_{t1} - \mathbf{k}_{b1}\\
            \mathbf{k}_{t2} - \mathbf{k}_{b2}
        \end{array}
    \right).
\end{align}
To relate these beating vectors to the moir\'{e} reciprocal lattice, $\{\mathbf{k}_{M1},\mathbf{k}_{M2}\}$, we employ the established layer--moir\'{e} transformations~\cite{Ref_uni_bi_axial_strain_model_2016}:
\begin{align}\label{Eq_ktkb_kM}
    \left(
        \begin{array}{c}
            \mathbf{k}_{t1}\\
            \mathbf{k}_{t2}
        \end{array}
    \right)
    = T_{Mt}^{T}
    \left(
        \begin{array}{c}
            \mathbf{k}_{M1}\\
            \mathbf{k}_{M2}
        \end{array}
    \right)
    , \quad
    \left(
        \begin{array}{c}
            \mathbf{k}_{b1}\\
            \mathbf{k}_{b2}
        \end{array}
    \right)
    = T_{Mb}^{T}
    \left(
        \begin{array}{c}
            \mathbf{k}_{M1}\\
            \mathbf{k}_{M2}
        \end{array}
    \right).
\end{align}
Substituting Eq.~\eqref{Eq_ktkb_kM} into the definition in Eq.~\eqref{Eq_Beating_def_fundamental} directly yields the generalized beating--moir\'{e} transformation:
\begin{align}\label{Eq_Beating_def}
    \left(
        \begin{array}{c}
            \mathbf{k}_{B1}\\
            \mathbf{k}_{B2}
        \end{array}
    \right)
    =
    \left(
        \begin{array}{cc}
            i-m & k-q \\
            j-n & l-r
        \end{array}
    \right)
    \left(
        \begin{array}{c}
            \mathbf{k}_{M1}\\
            \mathbf{k}_{M2}
        \end{array}
    \right).
\end{align}

This generalized transformation contrasts with more restrictive models~\cite{Ref_strain_PRL_2018, Ref_uni_bi_axial_strain_model_2016} that assume a simplified diagonal form:
\begin{align}\label{Eq_Beating_Def_2016}
    \left(
        \begin{array}{c}
            \mathbf{k}_{B1}\\
            \mathbf{k}_{B2}
        \end{array}
    \right)
    \equiv
    \left(
        \begin{array}{c}
            \mathbf{k}_{t1} - \mathbf{k}_{b1}\\
            \mathbf{k}_{t2} - \mathbf{k}_{b2}
        \end{array}
    \right)
    =
    \left(
        \begin{array}{cc}
            N_1 & 0 \\
            0 & N_2
        \end{array}
    \right)
    \left(
        \begin{array}{c}
            \mathbf{k}_{M1}\\
            \mathbf{k}_{M2}
        \end{array}
    \right).
\end{align}
This diagonal form enforces an alignment ($\mathbf{k}_{B\alpha}\parallel\mathbf{k}_{M\alpha}$), which corresponds to the special case of our general relation [Eq.~\eqref{Eq_Beating_def}] where the off-diagonal entries of the transformation matrix vanish. In the general case, no such alignment assumption is required.

\medskip
%\noindent\textbf{Real-space moiré--beating relation and the beating number.}
\noindent\textbf{Real-space Moir\'{e}--Beating Transformation and the Beating Number.}
The real-space relationship between the moir\'{e} and beating lattices is obtained by performing a Fourier transform on their reciprocal-space definition in Eq.~\eqref{Eq_Beating_def}. This yields the moir\'{e}--beating transformation:
\begin{align}\label{Eq_aMaB}
    \left(
        \begin{array}{c}
            \mathbf{a}_{M1}\\
            \mathbf{a}_{M2}
        \end{array}
    \right)
    = T_{MB}
    \left(
        \begin{array}{c}
            \mathbf{a}_{B1}\\
            \mathbf{a}_{B2}
        \end{array}
    \right),
\end{align}
with the transformation matrix $T_{MB}$ given by the difference between $T_{Mt}$ and $T_{Mb}$:
\begin{align}
    T_{MB}
    = T_{Mt}-T_{Mb}
     = \left(
        \begin{array}{cc}
            i-m & j-n\\
            k-q & l-r
        \end{array}
    \right). \notag
\end{align}
A direct consequence of this linear transformation is the relationship between the areas of the two unit cells:
\begin{align}
    \mathbf{a}_{M1}\times \mathbf{a}_{M2}
    = \det(T_{MB}) ~\mathbf{a}_{B1}\times \mathbf{a}_{B2}. \notag
\end{align}
This enables a rigorous definition of the beating number, $N_B$, as the determinant of $T_{MB}$:
\begin{align}\label{Eq_NB_def}
    N_B = \det(T_{MB}) = (i-m)(l-r) - (j-n)(k-q),
\end{align}
which quantifies the number of beating cells contained within a single moir\'{e} primitive cell.

This formalism also highlights the generality of our approach. Under the more restrictive diagonal convention shown in Eq.~\eqref{Eq_Beating_Def_2016}, our general definition for $N_B$ naturally reduces to the simple product $N_B = N_1 N_2$. This demonstrates that our definition of the beating number is a universal formula that correctly encompasses the special cases assumed in prior models.

\medskip
%\noindent\textbf{Real-space beating--layer relations.}
\noindent\textbf{Real-space Beating--Layer Transformation.}
In real space, the beating primitives can also be related to the layer primitives via the transformation matrices $T_{Bt}$ and $T_{Bb}$:
\begin{align}\label{Eq_aBatab}
    \left(
        \begin{array}{c}
            \mathbf{a}_{B1}\\
            \mathbf{a}_{B2}
        \end{array}
    \right)
    = T_{Bt} \left(
        \begin{array}{c}
            \mathbf{a}_{t1}\\
            \mathbf{a}_{t2}
        \end{array}
    \right)
    = T_{Bb} \left(
        \begin{array}{c}
            \mathbf{a}_{b1}\\
            \mathbf{a}_{b2}
        \end{array}
    \right).
\end{align}
These matrices can be derived by combining previously established relations, yielding $T_{Bt} = T_{MB}^{-1} T_{Mt}$ and $T_{Bb} = T_{MB}^{-1} T_{Mb}$. The explicit forms are:
\begin{align}
    & T_{Bt} = \frac{1}{N_B}
    \left(
        \begin{array}{cc}
            i(l-r)-k(j-n) & nl-jr\\
            iq-km         & l(i-m)-j(k-q)
        \end{array}
    \right), \nonumber \\
    & T_{Bb} = \frac{1}{N_B}
    \left(
        \begin{array}{cc}
            m(l-r)-q(j-n) & nl-jr\\
            iq-km         & r(i-m)-n(k-q)
        \end{array}
    \right). \nonumber
\end{align}
A key physical insight emerges when contrasting these matrices with their integer counterparts for the moir\'{e} lattice ($T_{Mt}, T_{Mb}$). While the moir\'{e} transformation matrices are strictly composed of integers, the beating transformation matrices ($T_{Bt}, T_{Bb}$) contain the prefactor $1/N_B$ and can therefore have fractional entries when $N_B\neq 1$. 
This fundamental mathematical difference proves that the beating pattern is, in general, not a true crystallographic superlattice of the bilayer. This has a direct consequence for the interpretation of real-space images: the spacings between adjacent bright maxima correspond to the beating periodicities, $(B_1,B_2,B_3) = \big(|\mathbf{a}_{B1}|,\,|\mathbf{a}_{B2}|,\,|\mathbf{a}_{B1}+\mathbf{a}_{B2}|\big)$, instead of those of the true moir\'{e} primitive cell, $(M_1,M_2,M_3) = \big(|\mathbf{a}_{M1}|,\,|\mathbf{a}_{M2}|,\,|\mathbf{a}_{M1}+\mathbf{a}_{M2}|\big)$ [Fig.~\ref{fig_01}(a)]. This highlights a critical point for accurate crystallographic decoding: while the beating periodicity is the quantity directly measured from real-space images, approximating it as the true moir\'{e} periodicity can propagate systematic errors into the determination of the fundamental transformation matrices and the overall geometric model.

\medskip
%\noindent\textbf{Reciprocal-space beating--layer relations and an observational consequence.}
\noindent\textbf{Reciprocal-space Beating--Layer Transformation.}
Finally, the layer reciprocal primitives can be expanded in terms of the beating reciprocal primitives via the transposed matrices $T_{Bt}^T$ and $T_{Bb}^T$:
\begin{align}\label{Eq_ktkb_kB}
    \left(
        \begin{array}{c}
            \mathbf{k}_{t1}\\
            \mathbf{k}_{t2}
        \end{array}
    \right)
    = T_{Bt}^{T} \left(
        \begin{array}{c}
            \mathbf{k}_{B1}\\
            \mathbf{k}_{B2}
        \end{array}
    \right), \quad
    \left(
        \begin{array}{c}
            \mathbf{k}_{b1}\\
            \mathbf{k}_{b2}
        \end{array}
    \right)
    = T_{Bb}^{T} \left(
        \begin{array}{c}
            \mathbf{k}_{B1}\\
            \mathbf{k}_{B2}
        \end{array}
    \right).
\end{align}
As established previously, the transformation matrices $T_{Bt}$ and $T_{Bb}$ can have fractional entries when $N_B\neq 1$. This implies that the layer diffraction peaks at $\{\mathbf{k}_{t1},\mathbf{k}_{t2}\}$ and $\{\mathbf{k}_{b1},\mathbf{k}_{b2}\}$ do not necessarily lie on the reciprocal lattice sites defined by integer combinations of $\{\mathbf{k}_{B1},\mathbf{k}_{B2}\}$. Conversely, they are always located exactly on the reciprocal lattice sites of the moir\'{e} superlattice, $\{\mathbf{k}_{M1},\mathbf{k}_{M2}\}$, as dictated by the integer-matrix transformations in Eq.~\eqref{Eq_ktkb_kM}. The general transformation in Eq.~\eqref{Eq_ktkb_kB}, which does not require the lattice alignment assumed in more restrictive models, provides the key mathematical relations for extracting the complete moir\'{e} geometry from the experimentally accessible data, as will be demonstrated in the subsequent sections.

\medskip
%\noindent\textbf{Complete parameter set.}
\noindent\textbf{The Moir\'{e} Crystallography Parameter Set.}
A complete description of the bilayer geometry combines the conventional twist and strain parameters $(\theta,\boldsymbol{\varepsilon})$ [Fig.~\ref{fig_01}(d)] with the crystallographic descriptors introduced here, namely the integer transformation matrices $(T_{Mt},T_{Mb})$ and the beating number $N_B$. We therefore use the parameter set $\{\theta,\boldsymbol{\varepsilon},(T_{Mt},T_{Mb}),N_B\}$ as the basis of our generalized moir\'{e} crystallography. The relations derived in this section provide the toolkit for determining these parameters from experimental data; the next section presents the step-by-step decoding algorithm.

\begin{figure*}[!t]
  \centering{
  \includegraphics[width=0.8\textwidth,trim=0.5mm 0.5mm 0.9mm 0.5mm,clip]{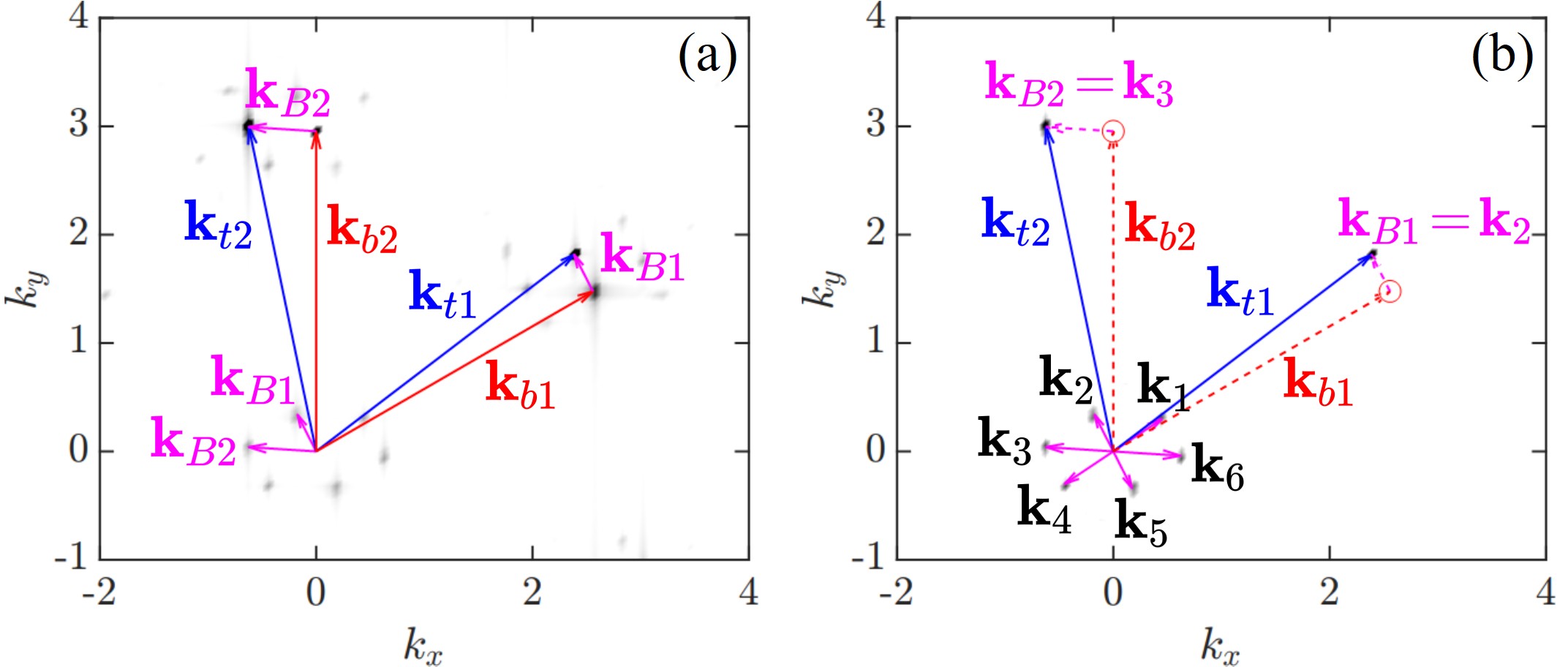}
  }
 \caption{
Reciprocal-space diffraction patterns for a bilayer specified by the integer transformation matrices $(i,j,k,l)=(5,-6,6,9)$ and $(m,n,q,r)=(6,-4,5,9)$, yielding a beating number of $N_B=2$. (a) Full detection: diffraction peaks from the top layer, bottom layer, and the beating pattern. Arrows indicate the corresponding primitive reciprocal-lattice vectors $\{\mathbf{k}_{t1},\mathbf{k}_{t2}\}$ (blue), $\{\mathbf{k}_{b1},\mathbf{k}_{b2}\}$ (red), and
$\{\mathbf{k}_{B1},\mathbf{k}_{B2}\}$ (magenta). (b) Bottom layer undetectable: only top-layer and beating peaks are observed. Red hollow circles mark a candidate reconstruction of the missing bottom-layer peaks obtained under the assumption that two observed beating peaks (e.g., $\mathbf{k}_2$ and $\mathbf{k}_3$) are taken to form the primitive beating basis $\{\mathbf{k}_{B1},\mathbf{k}_{B2}\}=\{\mathbf{k}_{2},\mathbf{k}_{3}\}$, together with  Eq.~\eqref{Eq_Beating_def_fundamental}.
 }
 \label{fig_Beating}
\end{figure*}

\begin{figure*}[t]
  \centering{
  \includegraphics[width=0.95\textwidth,trim=0.5mm 0.5mm 0.9mm 0.5mm,clip]{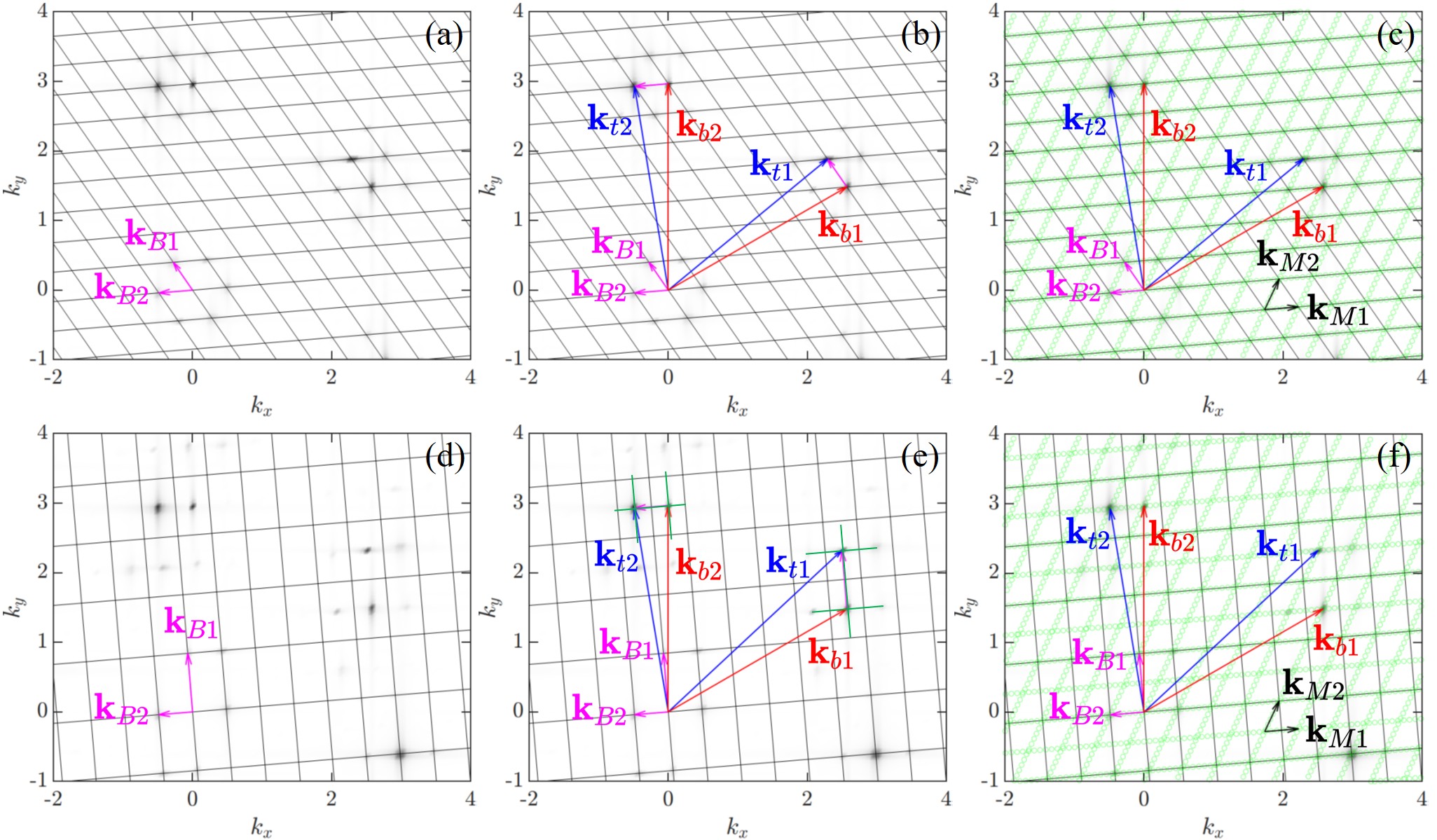}  
  }
 \caption{
 Schematic workflow for determining the integer transformation matrices from diffraction data, contrasting the special case $N_B=1$ (a–c) with the general case $N_B\neq1$ (d–f). Simulations use $(i,j,k,l,m,n,q,r)=(3,-4,4,7,\,4,-3,3,7)$ for $N_B=1$ and $(3,-4,5,7,\,4,-3,3,7)$ for $N_B=2$. (a,d) A reciprocal beating grid (black lines) is constructed from the primitive beating vectors $\{\mathbf{k}_{B1},\mathbf{k}_{B2}\}$ (magenta arrows). (b) For $N_B=1$, the top-layer ($\{\mathbf{k}_{t1},\mathbf{k}_{t2}\}$, blue arrows) and bottom-layer ($\{\mathbf{k}_{b1},\mathbf{k}_{b2}\}$, red arrows) peaks lie exactly on the beating-grid vertices, i.e., have integer coordinates. (c) The same peaks coincide with the vertices of the moir\'{e} reciprocal lattice (green hollow circles) defined by $\{\mathbf{k}_{M1},\mathbf{k}_{M2}\}$ (black arrows); in this special case the beating and moir\'{e} grids coincide. (e) For $N_B=2$, the layer peaks do not fall on beating-grid vertices; their fractional coordinates within each cell are obtained by subdividing the grid (green lines), yielding the coprime integers $(\alpha_{ij},\beta_{ij},N_{tb})$ of Eq.~\eqref{Eq_ktkb_kB_STM}. (f) In contrast, the layer peaks always land exactly on the moir\'{e} lattice vertices (green hollow circles), confirming that the moir\'{e} lattice is the Bravais lattice of the bilayer. These coordinates feed Eq.~\eqref{Eq_step4} to solve $(i,j,k,l,m,n,q,r)$.
 }
 \label{fig_NB_Eq1_Neq1}
\end{figure*}

\section{Decoding the Moir\'{e} Geometry from Imaging}\label{sec:Decoding}
% \subsection{Deriving Bottom Information from Top and Beating Data}
% \subsection{Reverse-Engineering Bottom-layer Information via Top-layer and Beating Data}

\subsection{Reconstructing the Buried-Layer Lattice}\label{sec:Step1}
We first address the experimentally common scenario in which the bottom layer is not atomically detectable. Our goal is to reconstruct the buried bottom-layer primitives from the observable top-layer and beating information. By the beating definition [Eq.~\eqref{Eq_Beating_def_fundamental}],
one can indirectly determine the bottom-layer vectors as \(\mathbf{k}_{b \alpha}=\mathbf{k}_{t  \alpha}-\mathbf{k}_{B  \alpha}\) ($ \alpha=1,2$) when \(\{\mathbf{k}_{t1},\mathbf{k}_{t2}\}\) and \(\{\mathbf{k}_{B1},\mathbf{k}_{B2}\}\) are known [Fig.~\ref{fig_Beating}(a)]. The practical challenge is a vector-assignment ambiguity: once \(\{\mathbf{k}_{t1},\mathbf{k}_{t2}\}\) are chosen from the six bright top-layer peaks away from \(\Gamma\) [Fig.~\ref{fig_Beating}(b)], the corresponding \(\{\mathbf{k}_{B1},\mathbf{k}_{B2}\}\) must be selected from the six beating peaks around \(\Gamma\), labeled \(\mathbf{k}_1,\ldots,\mathbf{k}_6\). Imposing that \(\mathbf{k}_{B1}\) and \(\mathbf{k}_{B2}\) are adjacent yields 12 candidate pairs:
\[
\{\mathbf{k}_{1}, \mathbf{k}_{2}\},\{\mathbf{k}_{2}, \mathbf{k}_{3}\},\ldots,\{\mathbf{k}_{6}, \mathbf{k}_{1}\},\;
\{\mathbf{k}_{2}, \mathbf{k}_{1}\},\{\mathbf{k}_{3}, \mathbf{k}_{2}\},\ldots,\{\mathbf{k}_{1}, \mathbf{k}_{6}\}.
\]
Each candidate \(\{\mathbf{k}_{B1},\mathbf{k}_{B2}\}\) produces a corresponding \(\{\mathbf{k}_{b1},\mathbf{k}_{b2}\}\).

To identify the physically correct pair from these candidates, we perform a self-consistency check for each one, leveraging both reciprocal- and real-space data through the following procedure:
\begin{enumerate}[label=\arabic*)]
    \item Generate a candidate $\{\mathbf{k}_{b1}^{\text{cand}}, \mathbf{k}_{b2}^{\text{cand}}\}$. For a given candidate pair \(\{\mathbf{k}_{B1}^{\text{cand}},\mathbf{k}_{B2}^{\text{cand}}\}\), assume it is the primitive beating basis and generate the corresponding candidate bottom-layer basis $\{\mathbf{k}_{b1}^{\text{cand}}, \mathbf{k}_{b2}^{\text{cand}}\}$ according to Eq.~\eqref{Eq_Beating_def_fundamental}.
    
    \item Solve for the moir\'{e}-layer transformation matrices $T_{Mt}$ and $T_{Mb}$. With the lattice vectors for both layers and beating now provisionally known, solve for the eight integer parameters $(i, j, k, l, m, n, q, r)$ that define the transformation matrices $T_{Mt}$ and $T_{Mb}$. The detailed solution method for this step is presented in Sec.~\ref{sec:Step2}.
    
    \item Calculate the corresponding beating spacings. Using the above solved $T_{Mt}$ and $T_{Mb}$, calculate the corresponding real-space beating spacings, $(B_1^{\text{cand}}, B_2^{\text{cand}}, B_3^{\text{cand}})$. This is done by first computing the matrix $T_{Bt}$ [\ Eq.~\eqref{Eq_aBatab}] and then using the measured top-layer lattice vectors $\{\mathbf{a}_{t1}, \mathbf{a}_{t2}\}$, the beating spacings follow a    
    \begin{align}
        & |\mathbf{a}_{B1}|=\sqrt{ T_{Bt,11}^{2}|\mathbf{a}_{t1}|^{2} + T_{Bt,12}^{2}|\mathbf{a}_{t2}|^{2} + 2(T_{Bt,11}T_{Bt,12})\mathbf{a}_{t1}\cdot\mathbf{a}_{t2}}, \nonumber\\
        & |\mathbf{a}_{B2}|=\sqrt{ T_{Bt,21}^{2}|\mathbf{a}_{t1}|^{2} + T_{Bt,22}^{2}|\mathbf{a}_{t2}|^{2} + 2(T_{Bt,21}T_{Bt,22})\mathbf{a}_{t1}\cdot\mathbf{a}_{t2}}, \nonumber\\
        & |\mathbf{a}_{B3}|=\sqrt{ T_{Bt,31}^{2}|\mathbf{a}_{t1}|^{2} + T_{Bt,32}^{2}|\mathbf{a}_{t2}|^{2} + 2(T_{Bt,31}T_{Bt,32})\mathbf{a}_{t1}\cdot\mathbf{a}_{t2}},
    \end{align}
    with $T_{Bt,31} = T_{Bt,11}+T_{Bt,21}$ and $T_{Bt,32} = T_{Bt,12}+T_{Bt,22}$. Here, $T_{Bt,ij}$ represents the $ij$-th matrix element of $T_{Bt}$, which is a function of the eight integers.
    
    \item Perform the consistency check. Compare the calculated beating spacings $(B_1^{\text{cand}}, B_2^{\text{cand}}, B_3^{\text{cand}})$ with the experimentally measured values $(B_1^{\text{exp}}, B_2^{\text{exp}}, B_3^{\text{exp}})$ obtained directly from real-space imaging.
\end{enumerate}

The candidate for which the calculated and experimental values are consistent identifies the physically relevant primitive beating lattice. This procedure thus reconstructs the buried-layer reciprocal vectors, $\{\mathbf{k}_{b1}, \mathbf{k}_{b2}\}$, solving the vector-assignment ambiguity. For the dataset in Figs.~\ref{fig_Beating}(a,b), only $\{\mathbf{k}_2,\mathbf{k}_3\}$ satisfies both the integer‑matrix constraints and the real‑space spacing consistency; the {\color[RGB]{0,0,255} Appendix A}  provides the step‑by‑step procedure and the full comparison across all 12 candidate pairs.

\subsection{Determining the Integer Transformation Matrices}\label{sec:Step2}

Once the lattice vectors for both layers, $\{\mathbf{k}_{t1}, \mathbf{k}_{t2}\}$ and $\{\mathbf{k}_{b1}, \mathbf{k}_{b2}\}$, are known (either from direct measurement or the reconstruction in Step 1), the central task is to determine the real-space integer transformation matrices, $T_{Mt}$ and $T_{Mb}$. This is achieved by solving for their eight integer components using parameters extracted from the reciprocal-space diffraction pattern. The procedure unfolds in two main stages, as visually demonstrated in Fig.~\ref{fig_NB_Eq1_Neq1}.

\medskip
\noindent First, parameterize the diffraction pattern with fractional coordinates. This initial stage converts the geometric information from the experimental diffraction pattern into a set of discrete integers. The procedure begins by establishing a ``reciprocal beating grid'' using the known primitive beating vectors, $\{\mathbf{k}_{B1}, \mathbf{k}_{B2}\}$, as a basis [Figs.~\ref{fig_NB_Eq1_Neq1}(a,d)]. A key visual distinction emerges at this stage: the layer peaks either coincide exactly with the grid vertices [Fig.~\ref{fig_NB_Eq1_Neq1}(b)], indicating an aligned system, or they fall at fractional coordinate positions within the grid cells [Fig.~\ref{fig_NB_Eq1_Neq1}(e)], signifying the general, non-aligned case.

For this general case, the fractional coordinates are determined by identifying the minimal grid subdivision required for all layer peaks to land on the new vertices (illustrated by the green lines in Fig.~\ref{fig_NB_Eq1_Neq1}(e)). The denominator of these fractions is the absolute value of the beating number, $N_B$, extracted directly from this reciprocal-space analysis. The coordinates are then expressed in fractional form:
\begin{align}\label{Eq_ktkb_kB_STM}
    \left(
        \begin{array}{c}
            \mathbf{k}_{t1}\\
            \mathbf{k}_{t2}
        \end{array}
    \right)
    = \frac{1}{N_B}
    \left(
        \begin{array}{cc}
            \alpha_{11} & \alpha_{12} \\
            \alpha_{21} & \alpha_{22}
        \end{array}
    \right)
    \left(
        \begin{array}{c}
            \mathbf{k}_{B1}\\
            \mathbf{k}_{B2}
        \end{array}
    \right), \\
    \left(
        \begin{array}{c}
            \mathbf{k}_{b1}\\
            \mathbf{k}_{b2}
        \end{array}
    \right)
    = \frac{1}{N_B}
    \left(
        \begin{array}{cc}
            \beta_{11} & \beta_{12} \\
            \beta_{21} & \beta_{22}
        \end{array}
    \right)
    \left(
        \begin{array}{c}
            \mathbf{k}_{B1}\\
            \mathbf{k}_{B2}
        \end{array}
    \right). \nonumber
\end{align}
The output is nine integers, specifically the numerators $\alpha_{ij}$, $\beta_{ij}$ and the common denominator $N_B$ , which are reduced to a setwise coprime form. For example, analysis of the pattern shown in Fig.~\ref{fig_NB_Eq1_Neq1}(e) yields $N_B=2$, $(\alpha_{11},\alpha_{12},\alpha_{21},\alpha_{22})=(5,-11,7,1)$, and $(\beta_{11},\beta_{12},\beta_{21},\beta_{22})=(3,-11,7,-1)$.

\medskip
\noindent Second, solve for the transformation matrices $T_{Mt}$ and $T_{Mb}$. 
%Equating Eq.~\eqref{Eq_ktkb_kB_STM} with the theoretical forms (Sec.~\ref{sec:framework}) yields the Diophantine system
Equating the experimental calibration [Eq.~\eqref{Eq_ktkb_kB_STM}] and the theoretical model (Eqs.~\eqref{Eq_NB_def} and \eqref{Eq_ktkb_kB} from Sec.~\ref{sec:framework}) yields a Diophantine system
\begin{align}\label{Eq_step4}
    \left(
        \begin{array}{cc}
            i(l-r)-k(j-n) & iq-km \\
            nl-jr         & l(i-m)-j(k-q)
        \end{array}
    \right)
    &=
    \left(
        \begin{array}{cc}
            \alpha_{11} & \alpha_{21} \\
            \alpha_{12} & \alpha_{22}
        \end{array}
    \right), \nonumber\\
    \left(
        \begin{array}{cc}
            m(l-r)-q(j-n) & iq-km \\
            nl-jr         & r(i-m)-n(k-q)
        \end{array}
    \right)
    &=
    \left(
        \begin{array}{cc}
            \beta_{11} & \beta_{21} \\
            \beta_{12} & \beta_{22}
        \end{array}
    \right), \nonumber\\
    (i-m)(l-r)-(j-n)(k-q)  &= N_B.
\end{align}
%From $\mathbf{k}_t-\mathbf{k}_b=\mathbf{k}_B$ one further obtains the consistency constraints $\alpha_{12}=\beta_{12}$, $\alpha_{21}=\beta_{21}$, and $\alpha_{11}-\beta_{11}=\alpha_{22}-\beta_{22}=N_B$, so there are seven independent equations for eight unknown integers, and multiple candidate solutions may exist. 
Note that in Eq.~\eqref{Eq_step4}, $nl-jr=\alpha_{12}=\beta_{12}$ and $iq-km=\alpha_{21}=\beta_{21}$, so there are seven independent equations for eight unknown integers, and multiple candidate solutions may exist. We bound the search by
\begin{align}\label{Eq_Aux}
    |i-m| \le \eta_{max}, ~~~
    |j-n| \le \lambda_{max}, ~~~
    |k-q| \le \mu_{max}, ~~~
    |l-r| \le \nu_{max},
\end{align}
and then solved numerically (using Mathematica's inbuilt \textit{Solve} function). Finally, Figs.~\ref{fig_NB_Eq1_Neq1}(c,f) visually confirm a key prediction: regardless of alignment, the layer peaks lie exactly on the vertices of the moir\'{e} reciprocal lattice (the Bravais lattice), while the six innermost peaks originate from the beating lattice.

% \subsection{Tunable parameters of HTB}
% \subsection{Correction from Tensile-Only to Full Strain Formalism}
% \subsection{Comprehensive Strain Formulation: Diverse Strain Conditions}
% \subsection{Enhanced Tunable-Parameter Formulation for Diverse Strain Scenarios}

\subsection{Extracting Geometric Control Parameters: Rotation and Strain $(\theta_r, \varepsilon_b, \theta_u, \varepsilon_u)$}\label{sec:TwistStrain}

With the two transformation matrices $T_{Mt}$ and $T_{Mb}$ determined in Sec.~\ref{sec:Step2}, the next step extracts the experimentally controllable parameters: the interlayer rotation $\theta_r$ and the strain components $(\varepsilon_b,\theta_u,\varepsilon_u)$. This uses the Park-Madden matrix formalism~\cite{Ref_Park_Madden,Ref_strain_PRL_2018}, which relates the layer lattices via a matrix $\mathcal{M}$ fully defined by the eight integers:
\begin{align}\label{Eq_Park_Madden}
    \left(
        \begin{array}{c}
            \mathbf{a}_{t1}\\
            \mathbf{a}_{t2}
        \end{array}
    \right)
    &=
    \frac{1}{il-jk}
    \left(
        \begin{array}{cc}
            lm-jq  &  nl-jr \\
            -km+iq &  -kn+ir
        \end{array}
    \right)
    \left(
        \begin{array}{c}
            \mathbf{a}_{b1}\\
            \mathbf{a}_{b2}
        \end{array}
    \right) \nonumber\\
    &\equiv
    \left(
        \begin{array}{cc}
            a & b \\
            c & d
        \end{array}
    \right)
    \left(
        \begin{array}{c}
            \mathbf{a}_{b1}\\
            \mathbf{a}_{b2}
        \end{array}
    \right)
    \equiv \mathcal{M}
    \left(
        \begin{array}{c}
            \mathbf{a}_{b1}\\
            \mathbf{a}_{b2}
        \end{array}
    \right).
\end{align}
The physical parameters $(\theta_r,\varepsilon_b,\theta_u,\varepsilon_u)$ are calculated from the elements $\{a,b,c,d\}$ using generalized analytical expressions derived in {\color{blue}Appendix~B}~[Eqs.~({\color{blue}B.10}),~({\color{blue}B.11}),~({\color{blue}B.18}),~({\color{blue}B.19})]. Our analysis incorporates two extensions relative to previous work~\cite{Ref_uni_bi_axial_strain_model_2016}: we explicitly include the Poisson's ratio ($\delta$) and rigorously treat the variable domains, allowing our expressions to capture both tensile and compressive strain solutions.

This generalized analysis reveals a tensile-compressive strain duality: each set of $(T_{Mt}, T_{Mb})$ maps to two distinct but related physical strain configurations. Specifically, for every solution corresponding to uniaxial tension $(\theta_{u}, \varepsilon_{u})$, a degenerate partner solution exists corresponding to uniaxial compression, related by the mapping:
\begin{equation}\label{Eq_TensileCompressiveMapping}
(\theta_{r},\varepsilon_b,\theta_u,\varepsilon_u)
\;\longleftrightarrow\;
\bigl(\theta_{r}, \varepsilon_{b}+(1-\delta)\varepsilon_{u},\; \theta_{u}\pm\frac{\pi}{2},\; -\varepsilon_{u}\bigr),
\end{equation}
where the sign in $\theta_{u}\pm\frac{\pi}{2}$ is chosen such that the angle remains in $[0,\pi)$. This two-branch structure originates from the intrinsic trigonometric periodicity of the analytical relations and reveals an inherent tensile–compressive duality in the moir\'{e} geometry that was not captured in previous analyses. In practice, the physically relevant branch can be selected using independent experimental constraints or energetic considerations from modeling.

\begin{figure*}[!t]
  \centering{
  \includegraphics[width=0.95\textwidth,trim=0.5mm 0.5mm 0.9mm 0.5mm,clip]{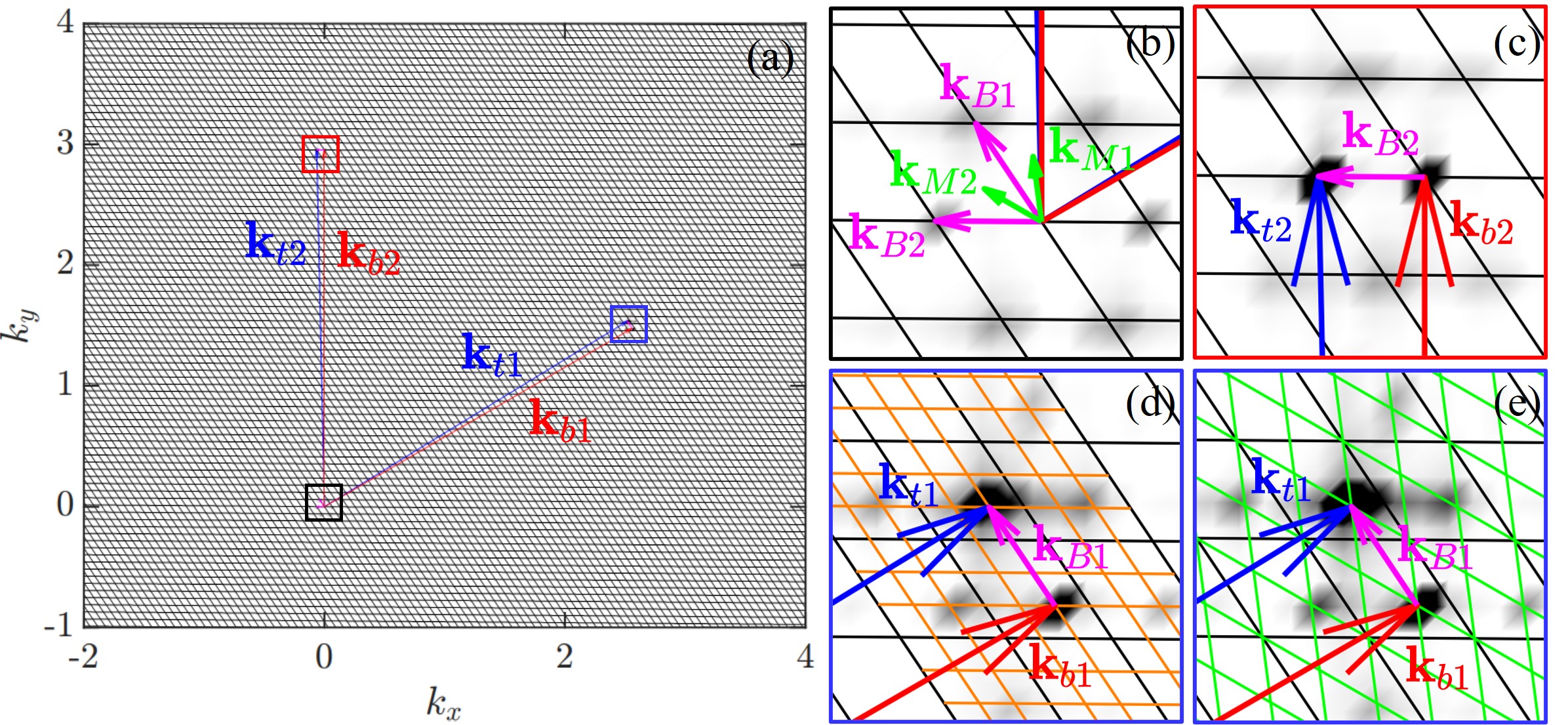}
  }
 \caption{
 Visualizing the moir\'{e} crystallography framework: a non-aligned case study. (a) Simulated diffraction pattern of the bilayer graphene system based on imaging data from Ref.~\cite{Ref_strain_PRL_2018}. The reciprocal beating grid (black lines) is defined by the observable beating vectors $\mathbf{k}_{B1}$ and $\mathbf{k}_{B2}$ (magenta arrows, shown magnified in (b)). (b-e) Magnified views of the regions marked by black, red, and blue boxes in (a). (b) Direct comparison between the beating basis $\{\mathbf{k}_{B1}, \mathbf{k}_{B2}\}$ (magenta) and the primitive moir\'{e} basis $\{\mathbf{k}_{M1}, \mathbf{k}_{M2}\}$ (green), showing their distinct orientations and magnitudes. (c) Integer occupancy on the beating grid. A specific direction where the layer diffraction peaks $\mathbf{k}_{t2}$ and $\mathbf{k}_{b2}$ (blue, red) coincide with the vertices of the beating grid, satisfying $\mathbf{k}_{t2} - \mathbf{k}_{b2} = \mathbf{k}_{B2}$. (d) Fractional occupancy and the moir\'{e} grid in the aligned ($\mathbf{k}_{B\alpha} \parallel \mathbf{k}_{M\alpha}$) configuration~\cite{Ref_strain_PRL_2018}. In contrast to (c), the layer peaks $\mathbf{k}_{t1}$ and $\mathbf{k}_{b1}$ are located at $1/3$ fractional positions relative to the beating grid. The orange sub-grid represents the moir\'{e} lattice derived by assuming the beating and moir\'{e} vectors are aligned~\cite{Ref_strain_PRL_2018}, which results in a redundant $N_B=9$ expansion. (e) Primitive moir\'{e} grid via general matrix decoupling. 
 %The green grid represents the result of our \textit{general matrix decoupling} method, which accounts for non-aligned configurations in the general case. 
 The green grid represents the result of our \textit{general matrix decoupling} method, which accounts for non-aligned configurations in the general case. 
 This primitive moir\'{e} grid identifies the $N_B=3$ Bravais lattice and coincides with all observed layer diffraction peaks $\{\mathbf{k}_{t1}, \mathbf{k}_{b1}, \mathbf{k}_{t2}, \mathbf{k}_{b2}\}$ without redundant expansion.
 }
 \label{fig_STM_to_ijklmnqr}
\end{figure*}

% \section{Results and Discussions}
%\section{Applications and Comparative Analysis}
\section{Case Study: Decoding a General Non-Aligned System}\label{sec:CaseStudy}

In this section, we re-examine the STM imaging data of the bilayer graphene system reported in Ref.~\cite{Ref_strain_PRL_2018}  by applying the decoding workflow and moir\'{e} crystallography framework established in the previous two sections. A central outcome is that this dataset corresponds to a general non-aligned geometry: indexing the data in the beating basis requires fractional coordinates, while a primitive moir\'{e} basis restores integer indexing. Consequently, the fundamental periodicity corresponds to a $N_B=3$ primitive moir\'{e} cell, whereas imposing an aligned (diagonal) constraint leads to a non-primitive $N_B=9$ supercell. This reassignment directly impacts the moir\'{e} Brillouin zone and the scale of \textit{ab initio} simulations.

Given the atomic thickness of both graphene layers, their diffraction signals are concurrently visible. The decoding workflow begins by partitioning the reciprocal space with a beating grid (black lines in Fig.~\ref{fig_STM_to_ijklmnqr}(a)) defined by the basis vectors $\{\mathbf{k}_{B1}, \mathbf{k}_{B2}\}$, which are denoted by the magenta arrows with an enlarged view provided in Fig.~\ref{fig_STM_to_ijklmnqr}(b). Through a direct grid-point count on the minimally subdivided beating grid within this reference frame, and following the parameterization in Eq.~\eqref{Eq_ktkb_kB_STM}, we map the layer diffraction peaks onto the beating basis to obtain:

\begin{align}\label{Eq_ktkb_kB_STM_2018}
    &
    \left(\!
       \begin{array}{ccccr}
         \mathbf{k}_{t1}\\
         \mathbf{k}_{t2}
       \end{array}\!
    \right)
    =
    \frac{1}{3}
    \left(
       \begin{array}{ccccr}
         79  \!&\! -167 \\
         150 \!&\! -90
       \end{array}
    \right)
    \left(\!
       \begin{array}{ccccr}
         \mathbf{k}_{B1}\\
         \mathbf{k}_{B2}
       \end{array}\!
    \right),\nonumber \\
   &
   \left(\!
       \begin{array}{ccccr}
         \mathbf{k}_{b1}\\
         \mathbf{k}_{b2}
       \end{array}\!
    \right)
    =
    \frac{1}{3}
    \left(\!
       \begin{array}{ccccr}
         76  \!&\! -167 \\
         150 \!&\! -93
       \end{array}\!
    \right)
    \left(\!
       \begin{array}{ccccr}
         \mathbf{k}_{B1}\\
         \mathbf{k}_{B2}
       \end{array}\!
    \right).
\end{align}

The common denominator directly identifies the beating number as $N_B = 3$. The extracted coefficients satisfy the consistency conditions described by Eq.~\eqref{Eq_ktkb_kB_STM}: here $\alpha_{12}=\beta_{12}=-167$, $\alpha_{21}=\beta_{21}=150$, and $\alpha_{11}-\beta_{11}=\alpha_{22}-\beta_{22}=N_B$ (e.g., $79-76=3$ and $(-90)-(-93)=3$), providing an internal check of the indexing. While the integer numerators align with the geometric configuration in Fig.~\ref{fig_STM_to_ijklmnqr}(c), the rational coordinates account for the fractional position signature observed in Fig.~\ref{fig_STM_to_ijklmnqr}(d). Together, these features indicate that the moir\'{e} and beating lattices are non-aligned, requiring the general (non-diagonal) matrix framework established in this work.

Substituting the fractional coordinates established in Eq.~\eqref{Eq_ktkb_kB_STM_2018} into the Diophantine system [Eq.~\eqref{Eq_step4}], and imposing the auxiliary search constraints from Eq.~\eqref{Eq_Aux} with $\eta_{\text{max}} = \lambda_{\text{max}} = \mu_{\text{max}} = \nu_{\text{max}} = 4$, we solve for the eight matrix elements $(i, j, k, l, m, n, q, r)$ of the real-space moir\'{e}–layer transformation matrices $(T_{Mt}, T_{Mb})$. As detailed in {\color{blue}Appendix~A}, we identify the physically consistent solution by comparing calculated beating point distances $(B_1^{\text{th}}, B_2^{\text{th}}, B_3^{\text{th}})$ with experimental STM measurements $(B_1^{\text{exp}}, B_2^{\text{exp}}, B_3^{\text{exp}})$ :
\begin{equation} \label{Eq_ijklmnqr}
(i, j, k, l, m, n, q, r) = (82, 80, -85, -10, 81, 81, -86, -12).
\end{equation}

These determined matrix elements define the reciprocal-space transformation between the beating and moir\'{e} bases according to Eq.~\eqref{Eq_Beating_def}:
\begin{equation} \label{Eq_kBi_kMi_thiswork}
\begin{pmatrix} \mathbf{k}_{B1} \\ \mathbf{k}_{B2} \end{pmatrix} = \begin{pmatrix} i-m & k-q \\ j-n & l-r \end{pmatrix} \begin{pmatrix} \mathbf{k}_{M1} \\ \mathbf{k}_{M2} \end{pmatrix} = \begin{pmatrix} 1 & 1 \\ -1 & 2 \end{pmatrix} \begin{pmatrix} \mathbf{k}_{M1} \\ \mathbf{k}_{M2} \end{pmatrix}.
\end{equation}
The non-diagonal form of this matrix represents the general case in which the beating and moir\'{e} basis vectors differ in both magnitude and orientation, as visualized in Fig.~\ref{fig_STM_to_ijklmnqr}(b). Its determinant is $\det \begin{pmatrix} 1 & 1 \\ -1 & 2 \end{pmatrix} = 3$, consistent with the denominator $N_B=3$ extracted in Eq.~\eqref{Eq_ktkb_kB_STM_2018}. This description contrasts with simplified models that require the two bases to be either identical ($\mathbf{k}_{B\alpha} = \mathbf{k}_{M\alpha}$) or strictly aligned ($\mathbf{k}_{B\alpha} \parallel \mathbf{k}_{M\alpha}$). Specifically, the model in Ref.~\cite{Ref_strain_PRL_2018} adopts the aligned assumption, yielding a diagonal transformation matrix: 
\begin{equation} \label{Eq_kBi_kMi_2018} 
\begin{pmatrix} \mathbf{k}_{B1} \\ \mathbf{k}_{B2} \end{pmatrix} = \begin{pmatrix} 3 & 0 \\ 0 & 3 \end{pmatrix} \begin{pmatrix} \mathbf{k}_{M1} \\ \mathbf{k}_{M2} \end{pmatrix}. 
\end{equation}
As visualized by the orange grid in Fig.~\ref{fig_STM_to_ijklmnqr}(d), the diagonal constraint in Eq.~(\ref{Eq_kBi_kMi_2018}) enforces a non-primitive $N_B = 9$ subdivision. In contrast, the non-diagonal transformation in Eq.~\eqref{Eq_kBi_kMi_thiswork} identifies the $N_B = 3$ primitive moir\'{e} lattice. As verified by the green grid in Fig.~\ref{fig_STM_to_ijklmnqr}(e), this mapping ensures that all top and bottom layer diffraction peaks land precisely on the integer vertices of the reconstructed moir\'{e} frame. This transition from the rational coordinates in Eq.~\eqref{Eq_ktkb_kB_STM_2018} to a purely integer-indexed moir\'{e} representation in Eq.~\eqref{Eq_ijklmnqr} confirms that the decoupling framework restores the fundamental periodicity of the non-aligned system. 

Utilizing the resolved integers $(i, j, k, l, m, n, q, r)$, the total number of atoms for the bilayer graphene system is calculated via $N_{\text{total}} = 2(|\det T_{Mt}| + |\det T_{Mb}|)$. For the $N_B=3$ primitive cell, we find $|\det T_{Mt}| = 5980$ and $|\det T_{Mb}| = 5994$, yielding a total of 23,948 atoms. In contrast, the non-primitive $N_B=9$ supercell reported in Ref.~\cite{Ref_strain_PRL_2018} necessitates 71,844 atoms to describe the same physical system, which matches the atom count reported in the literature. This substantial reduction in the atomic basis clarifies the system's minimal periodic scale and alleviates the computational complexity for \textit{ab initio} electronic structure simulations.

With the rigorous characterization of the beating--moir{\'e}, beating--layer, and moir{\'e}--layer transformation matrices, we determine the remaining structural descriptors $(\theta_r, \varepsilon_b, \theta_u, \varepsilon_u)$ that complete the moir{\'e} crystallography of this system. Substituting the resolved integers from Eq.~(\ref{Eq_ijklmnqr}) into the matrix formalism [Eq.~\eqref{Eq_Park_Madden}] and analytical expressions [Eq.~\eqref{Eq_TensileCompressiveMapping}] with a Poisson's ratio of $\delta = 0.16$ yields two candidate configurations, reflecting the tensile--compressive duality:
\begin{align}\label{Eq_physical_results}
    (\theta_r, \varepsilon_{b}, \theta_u, \varepsilon_{u}) = 
    \begin{cases} 
        (1.284^{\circ}, ~0.244\%, ~77.106^{\circ}, ~-0.301\%), \\ 
        (1.284^{\circ}, ~0.009\%, ~167.106^{\circ}, ~0.301\%). 
    \end{cases}
\end{align}
Both branches correspond to a twist angle of $\theta_r \approx 1.284^{\circ}$ and a uniaxial strain magnitude of $\varepsilon_u \approx 0.301\%$, with the physically relevant branch dictated by independent experimental constraints or energetic considerations.

At this stage, we have successfully resolved the complete set of moir{\'e} crystallographic parameters for the bilayer graphene system, encompassing $\{\theta_r, \varepsilon_b, \theta_u, \varepsilon_u, (T_{Mt}, T_{Mb}), N_B\}$. To highlight the significance of these results, we provide a comparative analysis between our generalized framework ($\mathbf{k}_{B\alpha}$ and $\mathbf{k}_{M\alpha}$ unconstrained) and the results derived from previous approximate models, specifically the aligned ($\mathbf{k}_{B\alpha} \parallel \mathbf{k}_{M\alpha}$)~\cite{Ref_uni_bi_axial_strain_model_2016, Ref_strain_PRL_2018} and identical ($\mathbf{k}_{B\alpha} = \mathbf{k}_{M\alpha}$)~\cite{Ref_Correlation_uniaxial_strain_model_2019,Ref_Correlation_uniaxial_strain_model_2020,Ref_strain_PRL_2021} configurations. These comparisons are summarized in Table~{\color{blue}A2} of {\color{blue}Appendix C}.

The results indicate that if the analysis is restricted to the determination of the four structural parameters $(\theta_r, \varepsilon_b, \theta_u, \varepsilon_u)$, conventional approximations may remain viable in the limit of small twist angles and weak strains. However, for applications where the primitive cell scale and the moir\'{e} Brillouin-zone construction are decisive, such as in the study of correlated electronic states and large-scale \textit{ab initio} simulations, the beating number $N_B$ and the integer transformation matrices play a fundamental role by fixing the minimal atomic basis and periodicity. Even under small twist angles and weak strain, conventional models can misassign these crystallographic descriptors in non-aligned geometries, motivating the generalized framework developed here for realistic moir{\'e} systems.

\section{Conclusion}

The precise geometric characterization of moir{\'e} bilayer superlattices from experimental imaging is foundational for understanding and engineering moir{\'e} quantum materials. In experiments, the most prominent long-period modulation inferred from real-space contrast and from the innermost satellite peaks often reflects the beating lattice, whereas the primitive moir{\'e} cell is defined as the minimal Bravais cell whose translations exactly repeat the full atomic configuration. These two notions coincide only in special situations. In general bilayers with arbitrary twist and strain states, the beating and moir{\'e} primitive vectors usually differ in both orientation and magnitude, and the buried layer can be difficult to resolve in systems with optically or electronically thick monolayers, which further restricts the information directly accessible from imaging.

In this work, we establish a generalized moir{\'e} crystallography framework by explicitly distinguishing the beating lattice from the moir{\'e} lattice and by formulating their exact transformation relations in both real and reciprocal space. The framework is organized around a unified hierarchy of three matrix classes: the moir{\'e}--layer matrices $(T_{Mt}, T_{Mb})$ defining the commensurate crystal, the beating--moir{\'e} transformation $(T_{MB})$ relating the visual and physical periodicities, and the beating--layer matrices $(T_{Bt}, T_{Bb})$ connecting experimental signals to the constituent lattices. This formulation encompasses commonly used prescriptions as special cases, where the identity and diagonal transformations correspond to the identity ($N_B=1$) and aligned limits, respectively. Our framework remains valid across arbitrary twist angles and general strain states, removing the constraints of small-angle and weak-strain approximations.

Building on this crystallographic formulation, we provide a decoding workflow that is compatible with experimentally constrained datasets via a combined analytical and numerical strategy. First, when the buried layer is not directly detectable, we reconstruct the missing layer lattice by utilizing the reciprocal-space beating definition together with a real-space consistency check. Second, we determine the integer transformation matrices by parameterizing diffraction peaks in the beating basis and solving the resulting Diophantine constraints to extract the beating number $N_B$ and the eight matrix elements. Third, we resolve the geometric control parameters $(\theta_r, \varepsilon_b, \theta_u, \varepsilon_u)$ from the Park--Madden matrix and the generalized analytical relations, which explicitly incorporate the Poisson effect and treat both tensile and compressive branches on equal footing. 

These results highlight that moir{\'e} geometry cannot be fully characterized by $(\theta_r, \varepsilon_b, \theta_u, \varepsilon_u)$ alone when the primitive cell scale and moir{\'e} Brillouin-zone construction are decisive. The beating number $N_B$ and the transformation matrices $(T_{Mt}, T_{Mb})$ are core descriptors that fix the minimal periodicity and the atomic basis, thereby controlling reciprocal-space indexing, band folding, and the length-scale competition central to correlated electronic phenomena. In our reanalysis of published STM data on twisted bilayer graphene, identifying the authentic $N_B=3$ primitive cell instead of a non-primitive $N_B=9$ supercell leads to a three-fold reduction in the atomic basis required for \textit{ab initio} simulations. By providing a complete and experimentally decodable descriptor set $\{\theta_r, \boldsymbol{\varepsilon}, (T_{Mt}, T_{Mb}), N_B\}$, our framework offers a consistent route from imaging data to quantitative atomistic models across a broad range of conditions, effectively bridging the gap between experimental characterization and the predictive design of realistic moir{\'e} quantum matter.

\section*{Author contributions}
X.H.K. conceived and supervised the project. Z.D.L. and X.H.K. developed the theory, analyzed the data, prepared the figures, and wrote the manuscript. Z.D.L. performed the analytical and numerical calculations. Both authors discussed the results and revised the manuscript.

\section*{Conflict of interest}
The authors declare no competing interests.

\section*{Acknowledgments}
We thank Prof. Hong Guo, Prof. Wei Ji, and Dr. Jiangbo Peng for helpful discussions. We gratefully acknowledge the financial support from the Shenzhen Science and Technology Innovation Commission under the Outstanding Youth Project (Grant No. RCYX20231211090126026), the National Natural Science Foundation of China (Grants No. 12474173, 52461160327), and Department of Science and Technology of Guangdong Province (Grants No. 2021QN02L820).

\bibliographystyle{elsarticle-num-names}
\bibliography{cas-refs}

\begin{thebibliography}{53}
\expandafter\ifx\csname natexlab\endcsname\relax\def\natexlab#1{#1}\fi
\providecommand{\url}[1]{\texttt{#1}}
\providecommand{\href}[2]{#2}
\providecommand{\path}[1]{#1}
\providecommand{\DOIprefix}{doi:}
\providecommand{\ArXivprefix}{arXiv:}
\providecommand{\URLprefix}{URL: }
\providecommand{\Pubmedprefix}{pmid:}
\providecommand{\doi}[1]{\href{http://dx.doi.org/#1}{\path{#1}}}
\providecommand{\Pubmed}[1]{\href{pmid:#1}{\path{#1}}}
\providecommand{\bibinfo}[2]{#2}
\ifx\xfnm\relax \def\xfnm[#1]{\unskip,\space#1}\fi
%Type = Article
\bibitem[{Lopes~dos Santos et~al.(2007)Lopes~dos Santos, Peres, and Castro~Neto}]{Ref_TBG_Theory_2007}
\bibinfo{author}{J.~M.~B. Lopes~dos Santos}, \bibinfo{author}{N.~M.~R. Peres}, \bibinfo{author}{A.~H. Castro~Neto},
\newblock \bibinfo{title}{Graphene bilayer with a twist: Electronic structure},
\newblock \bibinfo{journal}{Phys. Rev. Lett.} \bibinfo{volume}{99} (\bibinfo{year}{2007}) \bibinfo{pages}{256802}.
%Type = Article
\bibitem[{Su\'arez~Morell et~al.(2010)Su\'arez~Morell, Correa, Vargas, Pacheco, and Barticevic}]{Ref_TBG_Theory_2010_Tight_Binding_Model}
\bibinfo{author}{E.~Su\'arez~Morell}, \bibinfo{author}{J.~D. Correa}, \bibinfo{author}{P.~Vargas}, et~al.,
\newblock \bibinfo{title}{Flat bands in slightly twisted bilayer graphene: Tight-binding calculations},
\newblock \bibinfo{journal}{Phys. Rev. B} \bibinfo{volume}{82} (\bibinfo{year}{2010}) \bibinfo{pages}{121407}.
%Type = Article
\bibitem[{Trambly~de Laissardi\`ere et~al.(2012)Trambly~de Laissardi\`ere, Mayou, and Magaud}]{Ref_TBG_Theory_2012}
\bibinfo{author}{G.~Trambly~de Laissardi\`ere}, \bibinfo{author}{D.~Mayou}, \bibinfo{author}{L.~Magaud},
\newblock \bibinfo{title}{Numerical studies of confined states in rotated bilayers of graphene},
\newblock \bibinfo{journal}{Phys. Rev. B} \bibinfo{volume}{86} (\bibinfo{year}{2012}) \bibinfo{pages}{125413}.
%Type = Article
\bibitem[{Bistritzer and MacDonald(2011)}]{Ref_TBG_Theory_2011_BM_Model}
\bibinfo{author}{R.~Bistritzer}, \bibinfo{author}{A.~H. MacDonald},
\newblock \bibinfo{title}{Moiré bands in twisted double-layer graphene},
\newblock \bibinfo{journal}{Proc. Natl. Acad. Sci. U.S.A.} \bibinfo{volume}{108} (\bibinfo{year}{2011}) \bibinfo{pages}{12233--12237}.
%Type = Article
\bibitem[{Cao et~al.(2018{\natexlab{a}})Cao, Fatemi, Demir, Fang, Tomarken, Luo, Sanchez-Yamagishi, Watanabe, Taniguchi, Kaxiras, Ashoori, and Jarillo-Herrero}]{Ref_Cao_2018_Mott}
\bibinfo{author}{Y.~Cao}, \bibinfo{author}{V.~Fatemi}, \bibinfo{author}{A.~Demir}, et~al.,
\newblock \bibinfo{title}{Correlated insulator behaviour at half-filling in magic-angle graphene superlattices},
\newblock \bibinfo{journal}{Nature} \bibinfo{volume}{556} (\bibinfo{year}{2018}{\natexlab{a}}) \bibinfo{pages}{80--84}.
%Type = Article
\bibitem[{Cao et~al.(2018{\natexlab{b}})Cao, Fatemi, Fang, Watanabe, Taniguchi, Kaxiras, and Jarillo-Herrero}]{Ref_Cao_2018_SC}
\bibinfo{author}{Y.~Cao}, \bibinfo{author}{V.~Fatemi}, \bibinfo{author}{S.~Fang}, et~al.,
\newblock \bibinfo{title}{Unconventional superconductivity in magic-angle graphene superlattices},
\newblock \bibinfo{journal}{Nature} \bibinfo{volume}{556} (\bibinfo{year}{2018}{\natexlab{b}}) \bibinfo{pages}{43--50}.
%Type = Article
\bibitem[{Yankowitz et~al.(2019)Yankowitz, Chen, Polshyn, Zhang, Watanabe, Taniguchi, Graf, Young, and Dean}]{Ref_SC_Science_2019}
\bibinfo{author}{M.~Yankowitz}, \bibinfo{author}{S.~Chen}, \bibinfo{author}{H.~Polshyn}, et~al.,
\newblock \bibinfo{title}{Tuning superconductivity in twisted bilayer graphene},
\newblock \bibinfo{journal}{Science} \bibinfo{volume}{363} (\bibinfo{year}{2019}) \bibinfo{pages}{1059--1064}.
%Type = Article
\bibitem[{Xu and Balents(2018)}]{Ref_topo_SC_PRL_Balents_2018}
\bibinfo{author}{C.~Xu}, \bibinfo{author}{L.~Balents},
\newblock \bibinfo{title}{Topological superconductivity in twisted multilayer graphene},
\newblock \bibinfo{journal}{Phys. Rev. Lett.} \bibinfo{volume}{121} (\bibinfo{year}{2018}) \bibinfo{pages}{087001}.
%Type = Article
\bibitem[{Po et~al.(2018)Po, Zou, Vishwanath, and Senthil}]{Ref_Mott_SC_Senthil_PRX_2018}
\bibinfo{author}{H.~C. Po}, \bibinfo{author}{L.~Zou}, \bibinfo{author}{A.~Vishwanath}, et~al.,
\newblock \bibinfo{title}{Origin of mott insulating behavior and superconductivity in twisted bilayer graphene},
\newblock \bibinfo{journal}{Phys. Rev. X} \bibinfo{volume}{8} (\bibinfo{year}{2018}) \bibinfo{pages}{031089}.
%Type = Article
\bibitem[{Lee et~al.(2011)Lee, Riedl, Beringer, Castro~Neto, von Klitzing, Starke, and Smet}]{Ref_TBG_QHE_PRL_2011}
\bibinfo{author}{D.~S. Lee}, \bibinfo{author}{C.~Riedl}, \bibinfo{author}{T.~Beringer}, et~al.,
\newblock \bibinfo{title}{Quantum hall effect in twisted bilayer graphene},
\newblock \bibinfo{journal}{Phys. Rev. Lett.} \bibinfo{volume}{107} (\bibinfo{year}{2011}) \bibinfo{pages}{216602}.
%Type = Article
\bibitem[{Sanchez-Yamagishi et~al.(2012)Sanchez-Yamagishi, Taychatanapat, Watanabe, Taniguchi, Yacoby, and Jarillo-Herrero}]{Ref_TBG_QHE_PRL_2012}
\bibinfo{author}{J.~D. Sanchez-Yamagishi}, \bibinfo{author}{T.~Taychatanapat}, \bibinfo{author}{K.~Watanabe}, et~al.,
\newblock \bibinfo{title}{Quantum hall effect, screening, and layer-polarized insulating states in twisted bilayer graphene},
\newblock \bibinfo{journal}{Phys. Rev. Lett.} \bibinfo{volume}{108} (\bibinfo{year}{2012}) \bibinfo{pages}{076601}.
%Type = Article
\bibitem[{Moon and Koshino(2012)}]{Ref_TBG_QHE_PRB_2012}
\bibinfo{author}{P.~Moon}, \bibinfo{author}{M.~Koshino},
\newblock \bibinfo{title}{Energy spectrum and quantum hall effect in twisted bilayer graphene},
\newblock \bibinfo{journal}{Phys. Rev. B} \bibinfo{volume}{85} (\bibinfo{year}{2012}) \bibinfo{pages}{195458}.
%Type = Article
\bibitem[{Spanton et~al.(2018)Spanton, Zibrov, Zhou, Taniguchi, Watanabe, Zaletel, and Young}]{Ref_TBG_QHE_Science_2018}
\bibinfo{author}{E.~M. Spanton}, \bibinfo{author}{A.~A. Zibrov}, \bibinfo{author}{H.~Zhou}, et~al.,
\newblock \bibinfo{title}{Observation of fractional chern insulators in a van der waals heterostructure},
\newblock \bibinfo{journal}{Science} \bibinfo{volume}{360} (\bibinfo{year}{2018}) \bibinfo{pages}{62--66}.
%Type = Article
\bibitem[{Zhai et~al.(2023)Zhai, Chen, Xiao, and Yao}]{Ref_TBG_QHE_NC_2023}
\bibinfo{author}{D.~Zhai}, \bibinfo{author}{C.~Chen}, \bibinfo{author}{C.~Xiao}, et~al.,
\newblock \bibinfo{title}{Time-reversal even charge hall effect from twisted interface coupling},
\newblock \bibinfo{journal}{Nat. Commun.} \bibinfo{volume}{14} (\bibinfo{year}{2023}) \bibinfo{pages}{1961}.
%Type = Article
\bibitem[{Song et~al.(2019)Song, Wang, Shi, Li, Fang, and Bernevig}]{Ref_TBG_topology_PRL_Bernevig_2019}
\bibinfo{author}{Z.~Song}, \bibinfo{author}{Z.~Wang}, \bibinfo{author}{W.~Shi}, et~al.,
\newblock \bibinfo{title}{All magic angles in twisted bilayer graphene are topological},
\newblock \bibinfo{journal}{Phys. Rev. Lett.} \bibinfo{volume}{123} (\bibinfo{year}{2019}) \bibinfo{pages}{036401}.
%Type = Article
\bibitem[{Wu et~al.(2019)Wu, Lovorn, Tutuc, Martin, and MacDonald}]{Ref_TBG_topology_PRL_MacDonald_2019}
\bibinfo{author}{F.~Wu}, \bibinfo{author}{T.~Lovorn}, \bibinfo{author}{E.~Tutuc}, et~al.,
\newblock \bibinfo{title}{Topological insulators in twisted transition metal dichalcogenide homobilayers},
\newblock \bibinfo{journal}{Phys. Rev. Lett.} \bibinfo{volume}{122} (\bibinfo{year}{2019}) \bibinfo{pages}{086402}.
%Type = Article
\bibitem[{Hermann(2012)}]{Ref_Moiron_JPCM_2012}
\bibinfo{author}{K.~Hermann},
\newblock \bibinfo{title}{Periodic overlayers and moiré patterns: theoretical studies of geometric properties},
\newblock \bibinfo{journal}{J. Phys.: Condens. Matter} \bibinfo{volume}{24} (\bibinfo{year}{2012}) \bibinfo{pages}{314210}.
%Type = Article
\bibitem[{Iannuzzi et~al.(2013)Iannuzzi, Kalichava, Ma, Leake, Zhou, Li, Zhang, Bunk, Gao, Hutter, Willmott, and Greber}]{Ref_Moiron_PRB_2013}
\bibinfo{author}{M.~Iannuzzi}, \bibinfo{author}{I.~Kalichava}, \bibinfo{author}{H.~Ma}, et~al.,
\newblock \bibinfo{title}{Moir\'e beatings in graphene on ru(0001)},
\newblock \bibinfo{journal}{Phys. Rev. B} \bibinfo{volume}{88} (\bibinfo{year}{2013}) \bibinfo{pages}{125433}.
%Type = Article
\bibitem[{Alexeev et~al.(2019)Alexeev, Ruiz-Tijerina, Danovich, Hamer, Terry, Nayak, Ahn, Pak, Lee, Sohn, Molas, Koperski, Watanabe, Taniguchi, Novoselov, Gorbachev, Shin, Fal'ko, and Tartakovskii}]{Ref_Moiron_Nature_00_2019}
\bibinfo{author}{E.~M. Alexeev}, \bibinfo{author}{D.~A. Ruiz-Tijerina}, \bibinfo{author}{M.~Danovich}, et~al.,
\newblock \bibinfo{title}{Resonantly hybridized excitons in moir{\'e} superlattices in van der waals heterostructures},
\newblock \bibinfo{journal}{Nature} \bibinfo{volume}{567} (\bibinfo{year}{2019}) \bibinfo{pages}{81--86}.
%Type = Article
\bibitem[{Seyler et~al.(2019)Seyler, Rivera, Yu, Wilson, Ray, Mandrus, Yan, Yao, and Xu}]{Ref_Moiron_Nature_01_2019}
\bibinfo{author}{K.~L. Seyler}, \bibinfo{author}{P.~Rivera}, \bibinfo{author}{H.~Yu}, et~al.,
\newblock \bibinfo{title}{Signatures of moir{\'e}-trapped valley excitons in mose2/wse2 heterobilayers},
\newblock \bibinfo{journal}{Nature} \bibinfo{volume}{567} (\bibinfo{year}{2019}) \bibinfo{pages}{66--70}.
%Type = Article
\bibitem[{Tran et~al.(2019)Tran, Moody, Wu, Lu, Choi, Kim, Rai, Sanchez, Quan, Singh, Embley, Zepeda, Campbell, Autry, Taniguchi, Watanabe, Lu, Banerjee, Silverman, Kim, Tutuc, Yang, MacDonald, and Li}]{Ref_Moiron_Nature_02_2019}
\bibinfo{author}{K.~Tran}, \bibinfo{author}{G.~Moody}, \bibinfo{author}{F.~Wu}, et~al.,
\newblock \bibinfo{title}{Evidence for moir{\'e} excitons in van der waals heterostructures},
\newblock \bibinfo{journal}{Nature} \bibinfo{volume}{567} (\bibinfo{year}{2019}) \bibinfo{pages}{71--75}.
%Type = Article
\bibitem[{Cosma et~al.(2014)Cosma, Wallbank, Cheianov, and Fal{'}ko}]{Ref_strain_FaradayDiscuss_2014}
\bibinfo{author}{D.~A. Cosma}, \bibinfo{author}{J.~R. Wallbank}, \bibinfo{author}{V.~Cheianov}, et~al.,
\newblock \bibinfo{title}{Moiré pattern as a magnifying glass for strain and dislocations in van der waals heterostructures},
\newblock \bibinfo{journal}{Faraday Discuss.} \bibinfo{volume}{173} (\bibinfo{year}{2014}) \bibinfo{pages}{137--143}.
%Type = Article
\bibitem[{Bi et~al.(2019)Bi, Yuan, and Fu}]{Ref_strain_PRB_2019}
\bibinfo{author}{Z.~Bi}, \bibinfo{author}{N.~F.~Q. Yuan}, \bibinfo{author}{L.~Fu},
\newblock \bibinfo{title}{Designing flat bands by strain},
\newblock \bibinfo{journal}{Phys. Rev. B} \bibinfo{volume}{100} (\bibinfo{year}{2019}) \bibinfo{pages}{035448}.
%Type = Article
\bibitem[{Huder et~al.(2018)Huder, Artaud, Le~Quang, de~Laissardi\`ere, Jansen, Lapertot, Chapelier, and Renard}]{Ref_strain_PRL_2018}
\bibinfo{author}{L.~Huder}, \bibinfo{author}{A.~Artaud}, \bibinfo{author}{T.~Le~Quang}, et~al.,
\newblock \bibinfo{title}{Electronic spectrum of twisted graphene layers under heterostrain},
\newblock \bibinfo{journal}{Phys. Rev. Lett.} \bibinfo{volume}{120} (\bibinfo{year}{2018}) \bibinfo{pages}{156405}.
%Type = Article
\bibitem[{Mesple et~al.(2021)Mesple, Missaoui, Cea, Huder, Guinea, Trambly~de Laissardi\`ere, Chapelier, and Renard}]{Ref_strain_PRL_2021}
\bibinfo{author}{F.~Mesple}, \bibinfo{author}{A.~Missaoui}, \bibinfo{author}{T.~Cea}, et~al.,
\newblock \bibinfo{title}{Heterostrain determines flat bands in magic-angle twisted graphene layers},
\newblock \bibinfo{journal}{Phys. Rev. Lett.} \bibinfo{volume}{127} (\bibinfo{year}{2021}) \bibinfo{pages}{126405}.
%Type = Article
\bibitem[{Liu et~al.(2023)Liu, Zhou, Wu, and Yang}]{Ref_cupratemoire_2023_NC}
\bibinfo{author}{Y.-B. Liu}, \bibinfo{author}{J.~Zhou}, \bibinfo{author}{C.~Wu}, et~al.,
\newblock \bibinfo{title}{Charge-4e superconductivity and chiral metal in 45{\textdegree}-twisted bilayer cuprates and related bilayers},
\newblock \bibinfo{journal}{Nat. Commun.} \bibinfo{volume}{14} (\bibinfo{year}{2023}) \bibinfo{pages}{7926}.
%Type = Article
\bibitem[{Woods et~al.(2021)Woods, Ares, Nevison-Andrews, Holwill, Fabregas, Guinea, Geim, Novoselov, Walet, and Fumagalli}]{Ref_hBN_2021_NC}
\bibinfo{author}{C.~R. Woods}, \bibinfo{author}{P.~Ares}, \bibinfo{author}{H.~Nevison-Andrews}, et~al.,
\newblock \bibinfo{title}{Charge-polarized interfacial superlattices in marginally twisted hexagonal boron nitride},
\newblock \bibinfo{journal}{Nat. Commun.} \bibinfo{volume}{12} (\bibinfo{year}{2021}) \bibinfo{pages}{347}.
%Type = Article
\bibitem[{Kim et~al.(2024)Kim, Dominguez, Mayorga-Luna, Ye, Embley, Tan, Ni, Liu, Ford, Gao, Arash, Watanabe, Taniguchi, Kim, Shih, Lai, Yao, Yang, Li, and Miyahara}]{Ref_hBN_2024_NC}
\bibinfo{author}{D.~S. Kim}, \bibinfo{author}{R.~C. Dominguez}, \bibinfo{author}{R.~Mayorga-Luna}, et~al.,
\newblock \bibinfo{title}{Electrostatic moir{\'e} potential from twisted hexagonal boron nitride layers},
\newblock \bibinfo{journal}{Nat. Mater.} \bibinfo{volume}{23} (\bibinfo{year}{2024}) \bibinfo{pages}{65--70}.
%Type = Article
\bibitem[{Srivastava et~al.(2021)Srivastava, Hassan, de~Sousa, Gebredingle, Joe, Ali, Zheng, Yoo, Ghosh, Teherani, Singh, Low, and Lee}]{Ref_BlackP_2021_NE}
\bibinfo{author}{P.~K. Srivastava}, \bibinfo{author}{Y.~Hassan}, \bibinfo{author}{D.~J.~P. de~Sousa}, et~al.,
\newblock \bibinfo{title}{Resonant tunnelling diodes based on twisted black phosphorus homostructures},
\newblock \bibinfo{journal}{Nat. Electron.} \bibinfo{volume}{4} (\bibinfo{year}{2021}) \bibinfo{pages}{269--276}.
%Type = Article
\bibitem[{Chen et~al.(2024)Chen, Liang, Miao, Yu, Wang, Zhang, Wang, Wang, Cheng, Long, Wang, Wang, Zhang, and Chen}]{Ref_BlackP_2024_NC}
\bibinfo{author}{S.~Chen}, \bibinfo{author}{Z.~Liang}, \bibinfo{author}{J.~Miao}, et~al.,
\newblock \bibinfo{title}{Infrared optoelectronics in twisted black phosphorus},
\newblock \bibinfo{journal}{Nat. Commun.} \bibinfo{volume}{15} (\bibinfo{year}{2024}) \bibinfo{pages}{8834}.
%Type = Article
\bibitem[{Jin et~al.(2019)Jin, Regan, Yan, Iqbal Bakti~Utama, Wang, Zhao, Qin, Yang, Zheng, Shi, Watanabe, Taniguchi, Tongay, Zettl, and Wang}]{Ref_2019_TMDC_00}
\bibinfo{author}{C.~Jin}, \bibinfo{author}{E.~C. Regan}, \bibinfo{author}{A.~Yan}, et~al.,
\newblock \bibinfo{title}{Observation of moir{\'e} excitons in wse2/ws2 heterostructure superlattices},
\newblock \bibinfo{journal}{Nature} \bibinfo{volume}{567} (\bibinfo{year}{2019}) \bibinfo{pages}{76--80}.
%Type = Article
\bibitem[{Bai et~al.(2020)Bai, Zhou, Wang, Wu, McGilly, Halbertal, Lo, Liu, Ardelean, Rivera, Finney, Yang, Basov, Yao, Xu, Hone, Pasupathy, and Zhu}]{Ref_PFM_01}
\bibinfo{author}{Y.~Bai}, \bibinfo{author}{L.~Zhou}, \bibinfo{author}{J.~Wang}, et~al.,
\newblock \bibinfo{title}{Excitons in strain-induced one-dimensional moir{\'e} potentials at transition metal dichalcogenide heterojunctions},
\newblock \bibinfo{journal}{Nat. Mater.} \bibinfo{volume}{19} (\bibinfo{year}{2020}) \bibinfo{pages}{1068--1073}.
%Type = Article
\bibitem[{Liu et~al.(2025)Liu, Zhang, and Lu}]{Ref_exciton_2025_PNAS}
\bibinfo{author}{J.~Liu}, \bibinfo{author}{X.~Zhang}, \bibinfo{author}{G.~Lu},
\newblock \bibinfo{title}{Moiré magnetism and moiré excitons in twisted crsbr bilayers},
\newblock \bibinfo{journal}{Proc. Natl. Acad. Sci. U.S.A.} \bibinfo{volume}{122} (\bibinfo{year}{2025}) \bibinfo{pages}{e2413326121}.
%Type = Article
\bibitem[{Villafa\~ne et~al.(2023)Villafa\~ne, Kremser, H\"ubner, Petri\ifmmode~\acute{c}\else \'{c}\fi{}, Wilson, Stier, M\"uller, Florian, Steinhoff, and Finley}]{Ref_exciton_2023_PRL}
\bibinfo{author}{V.~Villafa\~ne}, \bibinfo{author}{M.~Kremser}, \bibinfo{author}{R.~H\"ubner}, et~al.,
\newblock \bibinfo{title}{Twist-dependent intra- and interlayer excitons in moir\'e ${\mathrm{mose}}_{2}$ homobilayers},
\newblock \bibinfo{journal}{Phys. Rev. Lett.} \bibinfo{volume}{130} (\bibinfo{year}{2023}) \bibinfo{pages}{026901}.
%Type = Article
\bibitem[{Mellado(2025)}]{Ref_MoireMagnet_00}
\bibinfo{author}{P.~Mellado},
\newblock \bibinfo{title}{Magnetic moiré systems: a review},
\newblock \bibinfo{journal}{J. Phys.:Condens. Matter} \bibinfo{volume}{37} (\bibinfo{year}{2025}) \bibinfo{pages}{323001}.
%Type = Article
\bibitem[{Gonzalez-Arraga et~al.(2017)Gonzalez-Arraga, Lado, Guinea, and San-Jose}]{Ref_MoireMagnet_01}
\bibinfo{author}{L.~A. Gonzalez-Arraga}, \bibinfo{author}{J.~L. Lado}, \bibinfo{author}{F.~Guinea}, et~al.,
\newblock \bibinfo{title}{Electrically controllable magnetism in twisted bilayer graphene},
\newblock \bibinfo{journal}{Phys. Rev. Lett.} \bibinfo{volume}{119} (\bibinfo{year}{2017}) \bibinfo{pages}{107201}.
%Type = Article
\bibitem[{Tao et~al.(2024)Tao, Shen, Jiang, Li, Li, Ma, Zhao, Hu, Pistunova, Watanabe, Taniguchi, Heinz, Mak, and Shan}]{Ref_MoireMagnet_02}
\bibinfo{author}{Z.~Tao}, \bibinfo{author}{B.~Shen}, \bibinfo{author}{S.~Jiang}, et~al.,
\newblock \bibinfo{title}{Valley-coherent quantum anomalous hall state in ab-stacked ${\mathrm{mote}}_{2}/{\mathrm{w}\mathrm{s}\mathrm{e}}_{2}$ bilayers},
\newblock \bibinfo{journal}{Phys. Rev. X} \bibinfo{volume}{14} (\bibinfo{year}{2024}) \bibinfo{pages}{011004}.
%Type = Article
\bibitem[{Li et~al.(2025)Li, Redekop, Wang~Beach, Zhang, Zhang, Liu, Holtzmann, Hu, Anderson, Park, Taniguchi, Watanabe, Chu, Fu, Cao, Xiao, Young, and Xu}]{Ref_MoireMagnet_03}
\bibinfo{author}{W.~Li}, \bibinfo{author}{E.~Redekop}, \bibinfo{author}{C.~Wang~Beach}, et~al.,
\newblock \bibinfo{title}{Universal magnetic phases in twisted bilayer mote2},
\newblock \bibinfo{journal}{Nano Lett.} \bibinfo{volume}{25} (\bibinfo{year}{2025}) \bibinfo{pages}{18044--18050}.
%Type = Article
\bibitem[{Artaud et~al.(2016)Artaud, Magaud, Le~Quang, Guisset, David, Chapelier, and Coraux}]{Ref_uni_bi_axial_strain_model_2016}
\bibinfo{author}{A.~Artaud}, \bibinfo{author}{L.~Magaud}, \bibinfo{author}{T.~Le~Quang}, et~al.,
\newblock \bibinfo{title}{Universal classification of twisted, strained and sheared graphene moir{\'e} superlattices},
\newblock \bibinfo{journal}{Sci. Rep.} \bibinfo{volume}{6} (\bibinfo{year}{2016}) \bibinfo{pages}{25670}.
%Type = Article
\bibitem[{Shabani et~al.(2021)Shabani, Halbertal, Wu, Chen, Liu, Hone, Yao, Basov, Zhu, and Pasupathy}]{Ref_STM_01}
\bibinfo{author}{S.~Shabani}, \bibinfo{author}{D.~Halbertal}, \bibinfo{author}{W.~Wu}, et~al.,
\newblock \bibinfo{title}{Deep moir{\'e} potentials in twisted transition metal dichalcogenide bilayers},
\newblock \bibinfo{journal}{Nat. Phys.} \bibinfo{volume}{17} (\bibinfo{year}{2021}) \bibinfo{pages}{720--725}.
%Type = Article
\bibitem[{Benschop et~al.(2021)Benschop, de~Jong, Stepanov, Lu, Stalman, van~der Molen, Efetov, and Allan}]{Ref_STM_02}
\bibinfo{author}{T.~Benschop}, \bibinfo{author}{T.~A. de~Jong}, \bibinfo{author}{P.~Stepanov}, et~al.,
\newblock \bibinfo{title}{Measuring local moir\'e lattice heterogeneity of twisted bilayer graphene},
\newblock \bibinfo{journal}{Phys. Rev. Res.} \bibinfo{volume}{3} (\bibinfo{year}{2021}) \bibinfo{pages}{013153}.
%Type = Article
\bibitem[{Woods et~al.(2014)Woods, Britnell, Eckmann, Ma, Lu, Guo, Lin, Yu, Cao, Gorbachev, Kretinin, Park, Ponomarenko, Katsnelson, Gornostyrev, Watanabe, Taniguchi, Casiraghi, Gao, Geim, and Novoselov}]{Ref_STM_03}
\bibinfo{author}{C.~R. Woods}, \bibinfo{author}{L.~Britnell}, \bibinfo{author}{A.~Eckmann}, et~al.,
\newblock \bibinfo{title}{Commensurate--incommensurate transition in graphene on hexagonal boron nitride},
\newblock \bibinfo{journal}{Nat. Phys.} \bibinfo{volume}{10} (\bibinfo{year}{2014}) \bibinfo{pages}{451--456}.
%Type = Article
\bibitem[{Jiang et~al.(2017)Jiang, Mao, Duan, Lai, Watanabe, Taniguchi, and Andrei}]{Ref_STM_04}
\bibinfo{author}{Y.~Jiang}, \bibinfo{author}{J.~Mao}, \bibinfo{author}{J.~Duan}, et~al.,
\newblock \bibinfo{title}{Visualizing strain-induced pseudomagnetic fields in graphene through an hbn magnifying glass},
\newblock \bibinfo{journal}{Nano Lett.} \bibinfo{volume}{17} (\bibinfo{year}{2017}) \bibinfo{pages}{2839--2843}.
%Type = Article
\bibitem[{Shi et~al.(2020)Shi, Zhan, Qi, Huang, Veen, Silva-Guill{\'e}n, Zhang, Li, Xie, Ji, Katsnelson, Yuan, Qin, and Zhang}]{Ref_STM_05}
\bibinfo{author}{H.~Shi}, \bibinfo{author}{Z.~Zhan}, \bibinfo{author}{Z.~Qi}, et~al.,
\newblock \bibinfo{title}{Large-area, periodic, and tunable intrinsic pseudo-magnetic fields in low-angle twisted bilayer graphene},
\newblock \bibinfo{journal}{Nat. Commun.} \bibinfo{volume}{11} (\bibinfo{year}{2020}) \bibinfo{pages}{371}.
%Type = Article
\bibitem[{Rosenberger et~al.(2020)Rosenberger, Chuang, Phillips, Oleshko, McCreary, Sivaram, Hellberg, and Jonker}]{Ref_cAFM_01}
\bibinfo{author}{M.~R. Rosenberger}, \bibinfo{author}{H.-J. Chuang}, \bibinfo{author}{M.~Phillips}, et~al.,
\newblock \bibinfo{title}{Twist angle-dependent atomic reconstruction and moiré patterns in transition metal dichalcogenide heterostructures},
\newblock \bibinfo{journal}{ACS Nano} \bibinfo{volume}{14} (\bibinfo{year}{2020}) \bibinfo{pages}{4550--4558}.
%Type = Article
\bibitem[{Zhang et~al.(2022)Zhang, Wang, Hu, Wan, Zheliuk, Liang, Peng, Zeng, and Ye}]{Ref_cAFM_02}
\bibinfo{author}{L.~Zhang}, \bibinfo{author}{Y.~Wang}, \bibinfo{author}{R.~Hu}, et~al.,
\newblock \bibinfo{title}{Correlated states in strained twisted bilayer graphenes away from the magic angle},
\newblock \bibinfo{journal}{Nano Lett.} \bibinfo{volume}{22} (\bibinfo{year}{2022}) \bibinfo{pages}{3204--3211}.
%Type = Article
\bibitem[{Yoo et~al.(2019)Yoo, Engelke, Carr, Fang, Zhang, Cazeaux, Sung, Hovden, Tsen, Taniguchi, Watanabe, Yi, Kim, Luskin, Tadmor, Kaxiras, and Kim}]{Ref_TEM_01}
\bibinfo{author}{H.~Yoo}, \bibinfo{author}{R.~Engelke}, \bibinfo{author}{S.~Carr}, et~al.,
\newblock \bibinfo{title}{Atomic and electronic reconstruction at the van der waals interface in twisted bilayer graphene},
\newblock \bibinfo{journal}{Nat. Mater.} \bibinfo{volume}{18} (\bibinfo{year}{2019}) \bibinfo{pages}{448--453}.
%Type = Article
\bibitem[{Alden et~al.(2013)Alden, Tsen, Huang, Hovden, Brown, Park, Muller, and McEuen}]{Ref_TEM_02}
\bibinfo{author}{J.~S. Alden}, \bibinfo{author}{A.~W. Tsen}, \bibinfo{author}{P.~Y. Huang}, et~al.,
\newblock \bibinfo{title}{Strain solitons and topological defects in bilayer graphene},
\newblock \bibinfo{journal}{Proc. Natl. Acad. Sci. U.S.A.} \bibinfo{volume}{110} (\bibinfo{year}{2013}) \bibinfo{pages}{11256--11260}.
%Type = Article
\bibitem[{Kazmierczak et~al.(2021)Kazmierczak, Van~Winkle, Ophus, Bustillo, Carr, Brown, Ciston, Taniguchi, Watanabe, and Bediako}]{Ref_TEM_03}
\bibinfo{author}{N.~P. Kazmierczak}, \bibinfo{author}{M.~Van~Winkle}, \bibinfo{author}{C.~Ophus}, et~al.,
\newblock \bibinfo{title}{Strain fields in twisted bilayer graphene},
\newblock \bibinfo{journal}{Nat. Mater.} \bibinfo{volume}{20} (\bibinfo{year}{2021}) \bibinfo{pages}{956--963}.
%Type = Article
\bibitem[{Weston et~al.(2020)Weston, Zou, Enaldiev, Summerfield, Clark, Z{\'o}lyomi, Graham, Yelgel, Magorrian, Zhou, Zultak, Hopkinson, Barinov, Bointon, Kretinin, Wilson, Beton, Fal'ko, Haigh, and Gorbachev}]{Ref_TEM_04}
\bibinfo{author}{A.~Weston}, \bibinfo{author}{Y.~Zou}, \bibinfo{author}{V.~Enaldiev}, et~al.,
\newblock \bibinfo{title}{Atomic reconstruction in twisted bilayers of transition metal dichalcogenides},
\newblock \bibinfo{journal}{Nat. Nanotechnol.} \bibinfo{volume}{15} (\bibinfo{year}{2020}) \bibinfo{pages}{592--597}.
%Type = Article
\bibitem[{Kerelsky et~al.(2019)Kerelsky, McGilly, Kennes, Xian, Yankowitz, Chen, Watanabe, Taniguchi, Hone, Dean, Rubio, and Pasupathy}]{Ref_Correlation_uniaxial_strain_model_2019}
\bibinfo{author}{A.~Kerelsky}, \bibinfo{author}{L.~J. McGilly}, \bibinfo{author}{D.~M. Kennes}, et~al.,
\newblock \bibinfo{title}{Maximized electron interactions at the magic angle in twisted bilayer graphene},
\newblock \bibinfo{journal}{Nature} \bibinfo{volume}{572} (\bibinfo{year}{2019}) \bibinfo{pages}{95--100}.
%Type = Article
\bibitem[{Zhang et~al.(2020)Zhang, Wang, Watanabe, Taniguchi, Ueno, Tutuc, and LeRoy}]{Ref_Correlation_uniaxial_strain_model_2020}
\bibinfo{author}{Z.~Zhang}, \bibinfo{author}{Y.~Wang}, \bibinfo{author}{K.~Watanabe}, et~al.,
\newblock \bibinfo{title}{Flat bands in twisted bilayer transition metal dichalcogenides},
\newblock \bibinfo{journal}{Nat. Phys.} \bibinfo{volume}{16} (\bibinfo{year}{2020}) \bibinfo{pages}{1093--1096}.
%Type = Article
\bibitem[{Park and Madden(1968)}]{Ref_Park_Madden}
\bibinfo{author}{R.~L. Park}, \bibinfo{author}{H.~H. Madden},
\newblock \bibinfo{title}{Annealing changes on the (100) surface of palladium and their effect on co adsorption},
\newblock \bibinfo{journal}{Surf. Sci.} \bibinfo{volume}{11} (\bibinfo{year}{1968}) \bibinfo{pages}{188--202}.

\end{thebibliography}

\end{document}